\documentclass[aps, prb, reprint, superscriptaddress]{revtex4-1}

\usepackage{amsmath}    
\usepackage{graphicx}   
\usepackage{verbatim}   
\usepackage{color}      
\usepackage{subfigure}  
\usepackage[colorlinks,linkcolor=blue,citecolor=blue,urlcolor=blue]{hyperref}
\usepackage{bbold}
\usepackage{amssymb}
\usepackage{float}

\newcommand\addtag{\refstepcounter{equation}\tag{\theequation}}

\newwrite\figurewrite
\immediate\openout\figurewrite=\jobname-figures.tex
\let\includegraphicsORI\includegraphics
\renewcommand{\includegraphics}[2][]{%
  \immediate\write\figurewrite{\unexpanded{\includegraphics[#1]{#2}}^^J}%
  \includegraphicsORI[#1]{#2}}
\newcommand{\myequation}{\begin{equation}}
\newcommand{\myendequation}{\end{equation}}
\let\[\myequation
\let\]\myendequation
\begin{document}

\title{Exploring excited eigenstates of many-body systems using the functional renormalization group}
\author{Christian Kl\"ockner}
\affiliation{Dahlem Center for Complex Quantum Systems and Fachbereich Physik, Freie Universit\"at Berlin, 14195 Berlin, Germany}
\email{kloeckner@physik.fu-berlin.de}
\author{Dante Marvin Kennes}
\affiliation{Department of Physics, Columbia University, New York, NY 10027, USA}
\author{Christoph Karrasch}
\affiliation{Dahlem Center for Complex Quantum Systems and Fachbereich Physik, Freie Universit\"at Berlin, 14195 Berlin, Germany}

\date{\today}

\begin{abstract}
We introduce approximate, functional renormalization group based schemes to obtain
correlation functions in pure excited eigenstates
of large fermionic many-body systems at arbitrary energies. The algorithms are
thoroughly benchmarked and their strengths and shortcomings are documented using a
one-dimensional interacting tight-binding chain as a prototypical testbed. We study
two `toy applications' from the world of Luttinger liquid physics: the survival of
power laws in lowly-excited states as well as the spectral function of high-energy `block' excitations which feature several single-particle Fermi edges.
\end{abstract}

\maketitle

\section{Introduction}

The complexity of diagonalizing the Hamiltonian or solving the time-dependent Schr\"odinger equation of generic interacting quantum problems grows exponentially with the number of particles involved. Over the last decades, a plethora of techniques were developed to study the physics of many-body systems which are in thermal equilibrium (most importantly in the ground state) as well as the out-of-equilibrium dynamics induced by pre-defined initial states. In contrast, far less attention was devoted to devising methods which can access pure excited eigenstates at arbitrary energies. This is, because up to very recently these questions were of minor relevance as pure excited eigenstates are difficult to realize in quantum many-body systems. However, the newly-emerging field of many-body localization,\cite{Gornyi2005,Basko2006} has changed this viewpoint drastically. This phenomenon, which cannot be analyzed using thermal ensembles as it defies our understanding of statistical mechanism as well as the eigenstate thermalization hypothesis~\cite{deutsch1991quantum,nandkishore2015many} renders the characterization and description of single excited eigenstates imperative. Before many-body localized systems were discovered it was believed that generic interactions wipe out the localization behavior induced by disorder in non-interacting systems~\cite{anderson1958absence} on the level of the eigenstates. To scrutinize this behavior one needs to go beyond the ground state as the prediction of many-body localization entails that the eigenstates of the entire spectrum can localize. Such a localization gives rise to interesting consequences for, e.g., the transport properties. Another fascinating question concerns the existence and characterization of a mobility edge separating localized and delocalized eigenstates in the spectrum. While in the non-interacting case such an edge does not appear for one-dimensional systems, first numerical results indicate that it will show up in interacting one-dimensional systems.~\cite{luitz2015many} To characterize this edge access to single eigenstates is essential.  Yet, a `golden standard'
approach to access single excited states has to be discovered. Exact diagonalization is restricted to small
systems. Proposals how to obtain a matrix product state representation of excited states are limited to one dimension and work only if the area law holds (i.e., in localized phases).~\cite{Khemani2015,Kennes2016,Friesdorf2015,Lim2016,Yu2017} This means that within this low-entanglement methods the crossover to the delocalization transition is difficult to approach. Analytical approaches~\cite{Moudgalya2017} such
as the Bethe ansatz can only be used for integrable models. Hence, it is desirable to
develop additional pure state techniques which feature their own, unique set of
strengths and shortcomings.

In this paper, we introduce several, functional renormalization group (fRG) based
algorithms to compute correlation functions in pure excited states of fermionic many-body systems. The fRG
extends standard Feynman perturbation theory by incorporating an RG idea on the
level of Green's functions.~\cite{Metzner2012,kopietzBook} The method was first set up in
thermal equilibrium (i.e., on the Matsubara axis) where it was used to study, e.g., instabilities in 2d systems\cite{halboth2000renormalization,honerkamp2001temperature,zanchi1998weakly} or the properties of quantum dots and wires.\cite{Andergassen2004,Enss2005,meden2006advances,bauer2014functional} Later on, the fRG was extended to Keldysh space and problems out of equilibrium were tackled~\cite{gezzi2007functional,Jakobs2007a,jakobs2010nonequilibrium,Karrasch2010,Kennes2012,Kennes2013,kennes2017adiabatically,sieberer2016keldysh} (see also Refs.~\onlinecite{salmhofer2001,katanin2004fulfillment,bartosch2009functional,reuther2010} for other developments). The key drawback of the method is the need to truncate its hierarchy of flow equations; thereafter, all results are only controlled up to a certain (usually first or
second) order in the two-particle interaction while still including a RG resummation of higher-order contributions. Thus, their validity needs to be
assessed carefully on a case-by-case basis. On the upside, the fRG can be used to access large
systems and is not a low-entanglement framework. Thus, an approximate, fRG-based
description of correlation functions in pure excited states would complement more accurate predictions
obtained, e.g., via exact diagonalization. It is the goal of this work to develop
and test such a `x-fRG' approach.

One way to obtain a pure eigenstate of a generic Hamiltonian is to analytically
determine an excited state of a noninteracting system and to then switch on
interactions adiabatically. We illustrate how to simplify the real-time Keldysh fRG
of Ref.~\onlinecite{Kennes2012} under the assumption of adiabaticity in order to
efficiently implement this protocol (this method will be called x-fRG-$t$-$\Gamma$).
Thereafter, a new RG cutoff scheme is devised which is specifically tailored to the
adiabatic nature of the problem (x-fRG-$t$-$\rho$). As an outlook and a potential route to go beyond linear order in the two-particle interactions, 
we briefly discuss how to obtain an
eigenstate directly (i.e., without resorting to the time evolution protocol) via a
coupling to a non-thermal bath (x-fRG-$\omega$) in Sec.~\ref{sec:xFRGw}.

We include an introduction to the fRG approach to both equilibrium and nonequilibrium problems and discuss some application specific aspects of these methods in lowest order truncation (Sec.~\ref{sec:matAndKeldFRG}), which form the basis for the new developments presented in this paper.
After developing the different x-fRG schemes and deriving the corresponding flow
equations (Sec.~\ref{ssec:exFrg}), we carry out several algorithmic tests and document the
limitations and promises of our approach (Sec.~\ref{sec:compMeth}). Thereafter, two `toy applications'
from the world of one-dimensional Luttinger liquid physics are presented: the
survival of power laws in lowly-excited states as well as the spectral function of
high-energy block excitations which feature multiple Fermi edges (Sec.~\ref{sec:appli}).

\section{Revisiting Matsubara and Keldysh fRG}
\label{sec:matAndKeldFRG}
In this section we will first introduce the models discussed in this paper as well as the notion of Green's functions. We then give a short introduction to the relevant aspects of the fRG formalism in Mastubara and Keldysh space which allow one to tackle equilibrium and time-dependent scenarios, respectively.
The x-fRG builds upon these concepts, and its derivation in Sec.~\ref{ssec:exFrg} will employ many of the relations that we discuss here. 
We also give a  detailed account of some algorithmic improvements of special significance for this work: a Trotter decomposition for Green's functions within time-dependent fRG (originally proposed in Ref.~\onlinecite{kennes2014universal}, see Sec.~\ref{ssec:tdfrg}) and details on how to evaluate relevant Green's functions in tight-binding models efficiently.
\subsection{Model}
We set up the fRG for general fermionic, particle-number conserving Hamiltonians with single- and two-particle contributions:
\[
    \begin{split}
        H&=H^0+H^\text{int}\\
        H^{0}=\sum_{i,j} h^0_{i,j} c_i^\dagger c_j &\hspace{0.5cm} H^\text{int}=\frac{1}{4}\sum_{i,j,k,l} u_{i,j,k,l} c_i^\dagger c_j^\dagger c_l c_k,
    \end{split}
\]
where \(c^{(\dagger)}_i\) are the fermionic annihilation (creation) operators and \(i\) denotes the single-particle index. \(H^0\) is the quadratic part of the Hamiltonian while \(H^\text{int}\) constitutes the two-particle interaction. The single-particle matrix representation of \(H^0\) is denoted by \(h^0\).

As an application, we will later on study one-dimensional tight-binding chains with \(N\) sites. We will therefore discuss how the final fRG flow equations can be solved efficiently for such models.
\subsection{Green's functions}
\subsubsection{Definitions}
An efficient way to describe nonequilibrium systems is the Keldysh formalism. Therein, the real-time single-particle correlation functions are defined as
\begin{alignat}{3}
    &\bigl[G^\text{ret}&&(t,t')\bigr]_{i,j}&&=-i\theta\left( t-t'\right) \left\langle \left[ c_i(t), c_j^\dagger(t')\right]_+\right\rangle_{\rho_0}\notag\\
    &G^\text{adv}&&(t,t')&&=\left[G^\text{ret}(t',t)\right]^\dagger\notag\\
    &\bigl[G^{<}&&(t,t')\bigr]_{i,j}&&=\hspace{0.25cm} i\left\langle c_j^\dagger(t') c_i(t)\right\rangle_{\rho_0}\\
    &\bigl[G^{>}&&(t,t')\bigr]_{i,j}&&=-i\left\langle c_i(t)c_j^\dagger(t') \right\rangle_{\rho_0}\notag\\
    &G^\text{K}&&(t,t')&&=G^<(t,t')+G^>(t,t')\notag
\end{alignat}
and are referred to as {\it retarded}, {\it advanced}, {\it lesser}, {\it greater} and {\it Keldysh} Green's function. The initial density matrix is given by
\[
    \addtag\label{eq:rho0}
    \begin{split}
        \rho_0&=\frac{1}{Z}\exp\left(-K\right),\hspace{0.5cm} Z=\text{Tr}\left[\exp\left(-K\right)\right]\\
        K&=\sum_{i,j}\bar K_{i,j} c_i^\dagger c_j,\hspace{0.5cm} \left\langle c_i^\dagger c_j\right\rangle_{\rho_0}=:\bar n_{j,i}=\left(\frac{\mathbb{1}}{e^{\bar{K}}+\mathbb{1}}\right)_{j,i},
    \end{split}
\] 
and \(\left\langle \dots\right\rangle_{\rho_0}=\text{Tr}\left(\rho_0\dots\right)\) denotes the corresponding expectation value. The time arguments are to be understood in the Heisenberg picture, and \([A,B]_{+}\) refers to the anticommutator. If not stated otherwise the initial time is set to \(t_0=0\).

In later chapters we will also consider the coupling to auxiliary reservoirs where we additionally assume the initial state to be a product-state of a quadratic density matrix in the system and the reservoirs.
\subsubsection{The noninteracting case}
If the system is described by a time-independent, noninteracting Hamiltonian \(H^0\), obtaining the Green's functions is straightforward. In the basis where \(H^0\) (but non necessarily \(\rho_0\)) is diagonal,
\[
    H^0=\sum_k\epsilon_k c_k^\dagger c_k,
\]
one  finds
\[
    \begin{split}
        \left[G^\text{ret}_0(t,t')\right]_{k,k^\prime}&=-i \theta(t-t')e^{-i \epsilon_{k}\left(t-t'\right)}\delta_{k,k'}\\
        \left[G^{<}_0(t,t')\right]_{k,k'}&=i e^{-i \epsilon_k t}\left\langle c_{k'}^\dagger c_k\right\rangle_{\rho_0} e^{i\epsilon_{k'}t'}
    \end{split}
\]
which transformed to the original basis yields
\[
    \addtag\label{eq:timeIndGF}
    \begin{split}
        G^\text{ret}_0(t,t')&=-i \theta(t-t')e^{-ih^0\left(t-t'\right)}\\
            G^{<}_0(t,t')  &=i e^{-i h^0 t} \bar n e^{i h^0t'}\\
        \Rightarrow
        G^\text{K}_0(t,t')  &=-iG^\text{ret}_0(t,0)\left(1-2\bar n\right)G^\text{adv}_0(0,t').
    \end{split}
\]
Here and in the following, matrix multiplications are implied in expressions such as \(G^\text{ret}_0(t,0) \bar n G^\text{adv}_0(0,t')\).

If \(H^0\) and \(\rho_0\)  commute (which is true, e.g., for a thermal state), the lesser and Keldysh Green's functions take a more recognizable form:
\[
    \addtag\label{eq:thermGtime}
    \begin{split}
        \left[ H^0, \rho_0\right]_-=0 \Rightarrow G^<_0(t,t')&=ie^{-ih^0(t-t')} \bar n,\\
        G^\text{K}_0(t,t')&=-ie^{-ih^0(t-t')}(1-2\bar n).
    \end{split}
\]

Analogously to Eq.~\eqref{eq:timeIndGF}, the case of a time-dependent noninteracting Hamiltonian can be treated:
\[
    \addtag \label{eq:gkFree}
    \begin{split}
    G^\text{ret}_{0}(t,t')&=-i\theta\left( t-t'\right) \mathcal{T} e^{-i\int_{t'}^{t} dt_1 h^0(t_1)}\\
    G^<_{0}(t,t')&=\ \ iG^\text{ret}_0(t,0)\bar n G^\text{adv}_0(0,t')\\
    G^\text{K}_{0}(t,t')&=-iG^\text{ret}_0(t,0) \left( 1-2\bar n\right)G^\text{adv}_0(0,t').
    \end{split}
\]
\subsubsection{The Dyson equation}
The two-particle interaction is taken into account via a self-energy.
The connection between the noninteracting and interacting Green's function is given by a Dyson equation:
\[
    G(t,t')=G_0(t,t')+\int_0^\infty dt_1 dt_2 G_0(t,t_1)\Sigma(t_1, t_2) G(t_2, t')
\]
\[
    G=
        \begin{pmatrix}
            G^\text{ret} & G^\text{K}\\
            0       & G^\text{adv}
        \end{pmatrix}
        \hspace{1cm}
    \Sigma=
        \begin{pmatrix}
            \Sigma^\text{ret} & \Sigma^\text{K}\\
            0       & \Sigma^\text{adv}
        \end{pmatrix}\addtag\label{eq:matDef}
\]
which for the individual components reads:
\[
    \addtag\label{eq:dysonTD}
    \begin{split}
        &G^\text{ret}(t,t')\\
        &=G^\text{ret}_0(t,t') +\int_0^\infty dt_1 dt_2 G^\text{ret}_0(t,t_1) \Sigma^\text{ret}(t_1,t_2) G^\text{ret}(t_2,t')\\
        &G^\text{K}(t,t')\\
        &=\left\{G^\text{ret}\left[\left(G^\text{ret}_0\right)^{-1}G^\text{K}_0\left(G^\text{adv}_0\right)^{-1}+\Sigma^\text{K}\right]G^\text{adv}\right\}(t,t')\\
                   &=-iG^\text{ret}(t,0)\left(1-2\bar n\right)G^\text{adv}(0, t')\\
                   &\hspace{0.5cm}+\int_0^\infty dt_1 dt_2 G^\text{ret}(t,t_1)\Sigma^\text{K}(t_1, t_2) G^\text{adv}(t_2, t'),
    \end{split}
\]
where we employed Eq.~\eqref{eq:gkFree}.
We will later on show that a leading-order fRG scheme yields self-energies which are time-local and have a vanishing Keldysh component:
\[
    \Sigma^\text{K}(t,t')=0,\hspace{1cm}\Sigma^\text{ret}(t,t')=\Sigma^\text{ret}(t) \delta(t-t'),
\]
which renders it particularly simple to evaluate Eq.~\eqref{eq:dysonTD}:

\[
    \begin{split}
         &G^\text{ret}(t,t')=-i\theta\left( t-t'\right) \mathcal{T} e^{-i\int_{t'}^{t} dt_1 \left[h^0(t_1)+\Sigma^\text{ret}(t_1)\right]}\\
        &G^\text{K}(t,t')=-iG^\text{ret}(t,0) \left( 1-2\bar n\right)G^\text{adv}(0,t').
    \end{split}
    \addtag\label{eq:simpelSelf}
\]
Hence, \(h^0+\Sigma^\text{ret}\) takes the role of an effective Hamiltonian.
Furthermore, the following group property holds in this situation:\cite{Kennes2012}
\begin{alignat}{1}
    \label{eq:groupProp}
    \text{for}\ t_i<\bar t<t_f:\notag\\
    \mathcal{T}e^{\left(\int_{t_i}^{t_f} dt\hat f(t)\right)}&=\mathcal{T}e^{\left(\int_{\bar t}^{t_f} dt\hat f(t)\right)}\mathcal{T}e^{\left(\int_{t_i}^{\bar t} dt\hat f(t)\right)}\notag\\
    \Rightarrow\ -iG^\text{ret}(t_i,t_f)&=G^\text{ret}(t_i,\bar t)G^\text{ret}(\bar t,t_f),
\end{alignat}
 which will allow for a major simplification of the fRG flow equations.
\subsubsection{Green's functions: steady state and equilibrium}
\label{sssec:gsinEq}
If the system reaches a steady state, all correlation functions in this state only depend on the time differences. Thus, one can define the Fourier transform as 
\[
    \begin{split}
        G^\text{ret}(\omega)&=\int_{-\infty}^\infty dt e^{i\omega (t-t')} G^\text{ret}(t-t')=\left[G^\text{adv}(\omega)\right]^\dagger\\
        G^\text{K}(\omega)&=\int_{-\infty}^\infty dt e^{i\omega (t-t')} G^\text{K}(t-t')=-\left[G^\text{K}(\omega)\right]^\dagger,
    \end{split}
\]
where we sent the initial time to \(t_0\rightarrow-\infty\).
In the noninteracting, time-independent case, the retarded Green's function takes the standard form
\[
    \addtag\label{eq:freeGFinOmeg}
    \begin{split}
        G^\text{ret}_0(\omega)&=\frac{1}{\omega-h^0+i0^+}.
    \end{split}
\]
The Dyson equation (see Eq.~\eqref{eq:dysonTD}) in a steady state reads
\[
    \begin{split}
        G^\text{ret}(\omega)&=G^\text{ret}_0(\omega)+G^\text{ret}_0(\omega)\Sigma^\text{ret}(\omega)G^\text{ret}(\omega)\\
        G^\text{K}(\omega)&=G^\text{ret}(\omega)\bigl[G^\text{ret}_0(\omega)^{-1} G^\text{K}_0(\omega) G^\text{adv}_0(\omega)^{-1} \\
        &\hspace{3cm}+\Sigma^\text{K}(\omega)\bigr] G^\text{adv}(\omega)\\
        &=G^\text{ret}(\omega) \Sigma^\text{K}(\omega) G^\text{adv}(\omega),
    \end{split}
\]
where in the second line of the last equation decay processes are assumed to lead to a fading memory of the initial density \(\rho_0\) in Eq.~\eqref{eq:dysonTD}.

In a thermal state (where \(\rho_0\) is not necessarily of the form of Eq.~\eqref{eq:rho0}),
\[
    \rho_0=e^{-\beta H}/\text{Tr}\left(e^{-\beta H}\right),
\]
the fluctuation-dissipation theorem (FDT) holds:
\[
    \addtag\label{eq:flucDis}
    G^\text{K}(\omega)=\left[1-2n_F(\omega)\right]\left[G^\text{ret}(\omega)-G^\text{adv}(\omega)\right]
\]
with the Fermi distribution \(n_F\)
\[
    1-2n_F(\omega)=\tanh\left(\left(\omega-\mu\right)\frac{\beta}{2}\right)
\]
at inverse temperature \(\beta\) and chemical potential \(\mu\).
Hence, all single-particle correlation functions can be extracted from \(G^\text{ret}\).

\subsubsection{Analytic continuation}
\label{sssec:anaCont}
We now elaborate how to obtain the equal-time Keldysh Green's function of an effectively noninteracting, time-independent system in thermal equilibrium efficiently using an analytic continuation; this will be important for later applications where we apply the same techniques for excited eigenstates.
Using the inverse Fourier transform, we find
\[
    G^\text{K}_0(t,t)=\frac{1}{2\pi}\int d\omega \left[1-2n_F(\omega)\right]\left[G^\text{ret}_0(\omega)-G^\text{adv}_0(\omega)\right].
\]
The time label \(t\) in equilibrium is arbitrary; we keep it to differentiate between time- and frequency-space. This expression can be recast as a contour integral,
\[
    \addtag \label{eq:contIntegi}
        G^\text{K}_0(t,t)=\frac{1}{2\pi}\sum_{\pm}\lim_{R\rightarrow \infty}\int_{\gamma^{R,1}_\pm} dz \left[1-2n_F(z)\right]G^\text{eq}_0(z),
\]
\begin{figure}
    \includegraphics[scale=0.7]{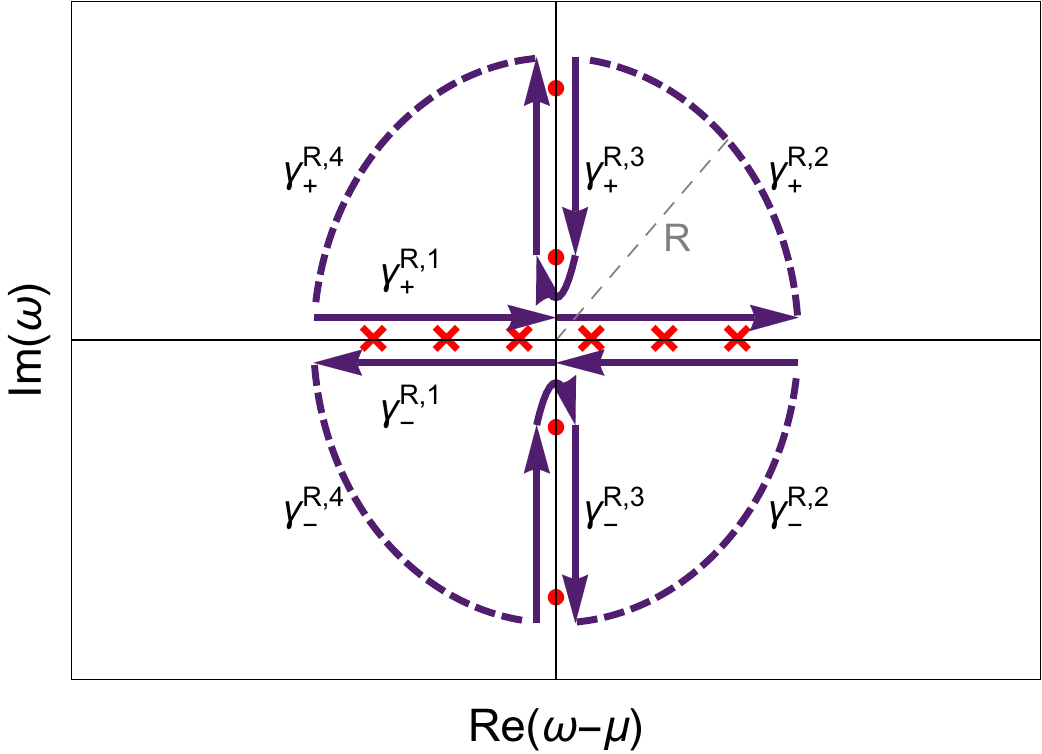}
    \caption{Schematic depiction of the integrations discussed in~\ref{sssec:anaCont}. The red crosses represent divergences at eigenenergies while the red circles signify divergences in the distribution function. The contours \(\gamma_\pm^{R,1}\) along the real axis are used to evaluate Eq.~\eqref{eq:contIntegi} while the contours \(\gamma_\pm^{R,3}\) enclosing the imaginary axis are used to evaluate Eq.~\eqref{eq:matSum}.}
   \label{fig:contour}
 \end{figure}%
where we introduced the equilibrium Green's function:
\[
    G^\text{eq}_0(z)=\frac{1}{z-h^0}
\]
\[
    \Rightarrow G^\text{ret}_0(\omega)=G^\text{eq}_0(\omega+i0^+),\hspace{0.5cm}G^\text{adv}_0(\omega)=G^\text{eq}_0(\omega-i0^+).
\]
The contours \(\gamma^{R,i}_\pm\), \(i=1,\dots,4\) are schematically shown in Fig.~\ref{fig:contour}.
The integrand of Eq.~\eqref{eq:contIntegi} has poles in the complex plane which are located at the eigenenergies of the Hamiltonian as well as at \(i\omega_n+\mu\) with the fermionic Matsubara frequencies \(i\omega_n\in i\frac{2\pi}{\beta}\left(\mathbb{Z}+\frac{1}{2}\right)\); the latter originate from the Fermi distribution. The integrand is analytic within the closed contours formed by \(\gamma^{R,1}_\pm+\dots+\gamma^{R,4}_\pm\) and at any finite chemical potential (and independently of the width enclosed by \(\gamma^{R,3}\))
\[
    \label{eq:curveContri}
    \begin{split}
        &\frac{1}{2\pi}\sum_{\pm}\lim_{R\rightarrow \infty}\int_{\gamma^{R,2}_\pm} dz \left[1-2n_F(z)\right]G^\text{eq}_0(z)\\
        =-&\frac{1}{2\pi}\sum_{\pm}\lim_{R\rightarrow \infty}\int_{\gamma^{R,4}_\pm} dz \left[1-2n_F(z)\right]G^\text{eq}_0(z),
    \end{split}
\]
as \(G^\text{eq}_0(z)\rightarrow 1/(z-\mu)+\mathcal{O}(1/(z-\mu)^2)\) for \(|z|\rightarrow\infty\).
Hence, the Keldysh Green's function is given by the contribution of the contour segment \(\gamma^{R,3}\).\footnote{The contribution by the contour element needed to close \(\gamma^{R,3}\) vanishes for \(R\rightarrow\infty\). It can always be chosen such, that the distance between the contour and the singularities remains finite.}
\[
    \addtag\label{eq:matSum}
    \begin{split}
    G^\text{K}_0(t,t)=&-\frac{1}{2\pi}\sum_{\pm}\lim_{R\rightarrow \infty}\int_{\gamma^{R,3}_\pm} dz \left[1-2n_F(z)\right]G^\text{eq}_0(z),\\
                     =&\frac{2i}{\beta}\lim_{N\rightarrow\infty}\sum_{n=-N}^N G^\text{eq}_0(i\omega_n+\mu)
    \end{split}
\]
and in the limit of \(T\rightarrow 0\)
\[
    \addtag\label{eq:gkT0}
    \begin{split}
        G^\text{K}_0(t,t) &=\frac{i}{\pi}\lim_{A\rightarrow\infty}\int_{-A}^A d\omega G^\text{eq}_0(i\omega+\mu)\\
        &=\frac{i}{\pi}\int d\omega e^{-|\omega|0^+}G^\text{eq}_0(i\omega+\mu).  
    \end{split}
\]
This can also be understood as the multipole expansion of the Fermi function.\cite{lin2009multipole}
One can show that for tight-binding models of linear dimension \(N\), the diagonal and first off-diagonal components of this expression can be evaluated in \(\mathcal{O}(N)\) operations.\cite{Andergassen2004,Usmani1994} This will be important for later applications where we want to study Luttinger liquid power-laws for \(N\gg 1\).
\subsection{General idea of fRG}
The functional renormalization group incorporates an RG idea on the level of correlation functions (for a review see, e.g., Ref.~\onlinecite{Metzner2012}). To this end, a low energy regularization is introduced in the noninteracting Green's function,
\[
    G_0\rightarrow G_0^\Lambda,
\]
which is chosen such that all vertex functions at the initial value of the cutoff parameter \(\Lambda\) are easy to evaluate and the physical Green's function is restored at its final value.
The fRG formalism yields an infinite set of differential equations which can be solved after a truncation. In this context, we restrict ourselves to the lowest order approximation, which only consists of a single equation of the schematic form:
\[
    \addtag\label{eq:generalFlow}
    \begin{split}
        \partial_\Lambda \Sigma^\Lambda&=uS^\Lambda\\
        S^\Lambda&=\partial_\Lambda^*G^\Lambda:=-G^\Lambda \left[ \partial_\Lambda \left(G^\Lambda_0\right)^{-1}\right]G^\Lambda.
    \end{split}
\]
\(S^\Lambda\) is referred to as single-scale propagator, which in the case that the cutoff enters as a self-energy (see, e.g. Sec.~\ref{ssec:keldFRG},~\ref{ssec:tdfrg}) simplifies to
\[
    \addtag\label{eq:singleScaleSelf}
    S^\Lambda=G^\Lambda \left(\partial_\Lambda \Sigma_\text{cut}^\Lambda\right) G^\Lambda.
\]
The precise form of Eq.~\eqref{eq:generalFlow} varies depending on the context and will be specified further in the later sections, where we also add the necessary indices and contractions. This scheme includes all contributions of first order as well as some terms of arbitrarily high order and in many cases regularizes infrared divergences.~\cite{Metzner2012}
\subsection{Matsubara fRG}
\label{ssec:gsfrg}
The simplest way to introduce a infrared regularization in Matsubara frequency space at \(T=0\) (i.e., in the ground state) is a sharp cutoff~\cite{Jakobs2007a} 
\[
    \addtag\label{eq:gsFlow}
    \begin{split}
        G^{\text{eq},\Lambda}_0(i\omega)&=G^\text{eq}_0(i\omega)\theta(|\omega|-\Lambda)\\
        S^{\text{eq},\Lambda}(i\omega)&=-\left[\delta(\omega-\Lambda)+\delta(\omega+\Lambda)\right]G^{\text{eq},\Lambda}(i\omega)\\
        G^{\text{eq},\Lambda}(i\omega)&=\frac{1}{i\omega-h^0-\Sigma^{\text{eq},\Lambda}}
    \end{split}
\]
which results in the following flow equation for the self-energy at \(T=0\):
\[
    \addtag\label{eq:flowMasFRG}
    \partial_\Lambda \Sigma_{i,j}^{\text{eq},\Lambda}=-\frac{1}{2\pi}\sum_{k,l}\sum_{\omega=\pm\Lambda} u_{i,k,j,l}\left(\frac{1}{i\omega-h^0-\Sigma^{\text{eq},\Lambda}}\right)_{l,k},
\]
where \(\Lambda\) is integrated from \(\infty\) to \(0\).
If one removes the feedback of the self-energy on the right-hand side of this equation, one obtains standard perturbation theory.
This illustrates that the fRG enhances perturbation theory by including a self-energy feedback; an integral is turned into a differential equation that treats different energy-scales successively in an RG sense.

\subsection{Keldysh fRG}
\label{ssec:keldFRG}

We will now briefly present the fRG framework in Keldysh space for steady states since this will serve as the basis for a generalization to excited states later. On the Keldysh axis, we employ a auxiliary reservoir cutoff scheme which is implemented by introducing a wide-band bath that couples to every single-particle degree of freedom of the system with a strength \(\Lambda\).~\cite{JakobsPHD}  Such a reservoir (which itself is in equilibrium) can be taken into account using a self-energy of the form
\[
    \addtag\label{eq:equiResSig}
    \begin{split}
        \Sigma^{\text{ret},\Lambda}_\text{cut}(\omega)&=-i\Lambda\mathbb{1}\\
        \Sigma^{\text{K},\Lambda}_\text{cut}(\omega)&=\left[1-2n_{\text{cut}}(\omega)\right]\underbrace{\left[\Sigma^{\text{ret},\Lambda}_\text{cut}(\omega)-\Sigma^{\text{adv},\Lambda}_\text{cut}(\omega)\right]}_{-2i\Lambda\mathbb{1}}
    \end{split}
\]
where \(n_{\text{cut}}(\omega)\) is its Fermi function and the second equation is enforced by the fluctuation-dissipation theorem. Thus, the reservoir-dressed Green's function is given by
\[
    \begin{split}
        G^{\text{ret}, \Lambda}(\omega)&=\frac{1}{G^\text{ret}_0(\omega)^{-1}+i\Lambda\mathbb{1}-\Sigma^{\text{ret},\Lambda}(\omega)}\\
        G^{\text{K}, \Lambda}(\omega)&=G^{\text{ret},\Lambda}(\omega)\left[ \Sigma^{\text{K},\Lambda}_\text{cut}(\omega)+\Sigma^{\text{K},\Lambda}(\omega)\right]G^{\text{adv},\Lambda}(\omega)\\
        &\stackrel{\text{FDT}}{=}\left[1-2n_\text{cut}(\omega)\right]\left[G^{\text{ret},\Lambda}(\omega)-G^{\text{adv},\Lambda}(\omega)\right]
    \end{split}
\]
and one can show (compare the retarded and Keldysh component of Eq.~\eqref{eq:singleScaleSelf}) that
\[
    \begin{split}
        S^{\text{ret},\Lambda}(\omega)&=\partial_\Lambda^* G^{\text{ret},\Lambda}(\omega)=-iG^{\text{ret},\Lambda}(\omega)G^{\text{ret},\Lambda}(\omega)\\
        S^{\text{K}, \Lambda}(\omega)&=\partial^*_\Lambda G^{\text{K},\Lambda}(\omega)\\
        &=S^{\text{ret},\Lambda}(\omega)\left[ \Sigma^{\text{K},\Lambda}_\text{cut}(\omega)+\Sigma^{\text{K},\Lambda}(\omega)\right]G^{\text{adv},\Lambda}(\omega)\\
                              &\hspace{0.2cm}+G^{\text{ret},\Lambda}(\omega)\left[ \Sigma^{\text{K},\Lambda}_\text{cut}(\omega)+\Sigma^{\text{K},\Lambda}(\omega)\right]S^{\text{adv},\Lambda}(\omega)\\
                              &\hspace{0.2cm}+G^{\text{ret},\Lambda}(\omega)\left[\partial_\Lambda \Sigma^{\text{K},\Lambda}_\text{cut}(\omega)\right]G^{\text{adv},\Lambda}(\omega)\\
        &\stackrel{\text{FDT}}{=}\left[1-2n_\text{cut}(\omega)\right]\left[S^{\text{ret},\Lambda}(\omega)-S^{\text{adv},\Lambda}(\omega)\right].
    \end{split}
\]
Here, the first lines contain the general steady-state expression; the equilibrium limit, which is described by a temperature \(T\geq 0\) and where the fluctuation-dissipation theorem holds, is denoted by FDT.
The flow equations read
\[
    \addtag\label{eq:flowKeldFRG}
    \begin{split}
        \partial_\Lambda \Sigma^{\text{ret},\Lambda}_{i,j}&=-\frac{i}{4\pi}\sum_{k,l} u_{i,k,j,l} \int d\omega \left[S^{\text{K},\Lambda}(\omega)\right]_{l,k}\\
        &\stackrel{\text{FDT}}{=}-\frac{i}{4\pi}\sum_{k,l} u_{i,k,j,l} \int d\omega \ \text{sgn}(\omega)\\
        &\hspace{1cm}\left[ \partial_\Lambda^*G^{\text{ret},\Lambda}(\omega)-\partial_\Lambda^*G^{\text{adv},\Lambda}(\omega)\right]_{l,k}\\
        &=-\frac{i}{4\pi}\sum_{k,l} u_{i,k,j,l} \int d\omega\ \text{sgn}(\omega)\\
        &\hspace{1cm}\left[ i\partial_\omega G^{\text{ret},\Lambda}(\omega)+i\partial_\omega G^{\text{adv},\Lambda}(\omega)\right]_{l,k}\\
        &=-\frac{1}{2\pi}\sum_{k,l} u_{i,k,j,l} \left[G^{\text{ret},\Lambda}(0)+G^{\text{adv},\Lambda}(0)\right]_{l,k}\\
    \end{split}
\]
\[
    \begin{split}
        \partial_\Lambda \Sigma^{\text{K},\Lambda}_{i,j}&=-\frac{i}{4\pi}\sum_{k,l} u_{i,k,j,l} \int d\omega \left[S^{\text{ret},\Lambda}(\omega)+S^{\text{adv},\Lambda}(\omega)\right]_{l,k}\\
        &=-\frac{i}{4\pi}\sum_{k,l} u_{i,k,j,l} \\
        &\hspace{1cm}\int d\omega \left[ i\partial_\omega G^{\text{ret},\Lambda}(\omega)-i\partial_\omega G^{\text{adv},\Lambda}(\omega)\right]_{l,k}\\
        &=0
    \end{split}
\]
where in equilibrium (denoted by FDT) we now focused on the \(T=0\) case; the final flow equations are identical to the ones obtained using the sharp Matsubara frequency cutoff (the chemical potential was assumed to be \(0\)).

Note that the Keldysh fRG was first developed for open systems,\cite{jakobs2010nonequilibrium,gezzi2007functional,Jakobs2007a,Karrasch2010} which can be treated in a similar fashion by adding an additional bath self-energy \(\Sigma^{\text{ret},\text{K}}_\text{lead}\). The structure of the flow equations does not change.
\subsection{Time-dependent fRG}
\label{ssec:tdfrg}
A time-dependent formulation of the fRG which employs the reservoir cutoff presented in Sec.~\ref{ssec:keldFRG} has been established in Ref.~\onlinecite{Kennes2012}.
In this scheme, all single-particle states of the system are initially coupled infinitely strongly to a wide-band reservoir which suppresses all dynamics and immediately forces the system into a thermal state regardless of the interaction; by lowering the hybridization strength \(\Lambda\), one eventually retrieves the original problem. If we conveniently choose the reservoir's temperature to be infinite, its effects can be understood in terms of the following self-energy (compare Eq.~\eqref{eq:equiResSig}):
\[
    \addtag\label{eq:resSelf}
    \begin{split}
        \Sigma^{\text{K},\Lambda}_\text{cut}(t,t')&=0\\
        \Sigma^{\text{ret},\Lambda}_\text{cut}(t,t')&=-i\delta(t-t') \Lambda\mathbb{1}.
    \end{split}
\]
The Green's functions of the time-dependent, noninteracting system read (compare Eq.~\eqref{eq:simpelSelf})
\begin{alignat}{2}
    \label{eq:resG0Decomp}
    &G^{\text{ret}, \Lambda}_0(t,t')&&=e^{-(t-t')\Lambda}G^{\text{ret},\Lambda=0}_0(t,t')\\
    &G^{\text{K},\Lambda}_{0}(t,t')&&=-iG^{\text{ret},\Lambda}_0(t,0) \left( 1-2\bar n\right)G^{\text{adv},\Lambda}_0(0,t').\notag
\end{alignat}
The fRG flow equations truncated to leading order are given by (a detailed derivation of this can be found in Ref.~\onlinecite{Kennes2012}):
\[
    \addtag\label{eq:genTflow}
    \begin{split}
        \partial_\Lambda \Sigma^{\text{ret},\Lambda}_{i,j}(t,t')&=-\frac{i}{2}\delta(t-t') \sum_{k,l} \left[S^{\text{K},\Lambda}(t,t)\right]_{l,k} u_{i,k,j,l}(t)\\
        \partial_\Lambda \Sigma^{\text{K},\Lambda}_{i,j}(t,t')&\\
        &\hspace{-1.5cm}=-\frac{i}{2}\delta(t-t') \sum_{k,l} \left(S^{\text{ret},\Lambda}+S^{\text{adv},\Lambda}\right)_{l,k}(t,t) u_{i,k,j,l}(t)\\
    \end{split}
\]
with the single-scale propagator (see Eq.~\eqref{eq:singleScaleSelf})
\[
    \addtag
    \label{eq:tdSS}
    \begin{split}
        &S^{\text{ret},\Lambda}(t,t')\\
        &=\int dt_1 dt_2 G^{\text{ret},\Lambda}(t,t_1) \left[\partial_\Lambda \Sigma^{\text{ret},\Lambda}_\text{cut}(t_1,t_2)\right] G^{\text{ret},\Lambda}(t_2,t')\\
        &S^{\text{K},\Lambda}(t,t')\\
                   &=\int dt_1 dt_2 \Bigl\{G^{\text{ret},\Lambda}(t,t_1) \left[\partial_\Lambda \Sigma^{\text{ret},\Lambda}_\text{cut}(t_1,t_2)\right] G^{\text{K},\Lambda}(t_2,t')\\
                    &\hspace{1cm}+G^{\text{ret},\Lambda}(t,t_1) \left[\partial_\Lambda \Sigma^{\text{K},\Lambda}_\text{cut}(t_1,t_2)\right] G^{\text{adv},\Lambda}(t_2,t')\\
                    &\hspace{1cm}+G^{\text{K},\Lambda}(t,t_1) \left[\partial_\Lambda \Sigma^{\text{adv},\Lambda}_\text{cut}(t_1,t_2)\right] G^{\text{adv},\Lambda}(t_2,t')\Bigr\},
    \end{split}
\]
where \(G^\Lambda\) refers to the full Green's function obtained via Eq.~\eqref{eq:dysonTD}. 
For the specific cutoff used here (see Eq.~\eqref{eq:resSelf}) the above equation takes the form
\[
    \addtag
    \label{eq:tdSSsimp}
    \begin{split}
        S^{\text{ret},\Lambda}(t,t')&=-i\int dt_1 G^{\text{ret},\Lambda}(t,t_1)G^{\text{ret},\Lambda}(t_1,t')\\
        \Rightarrow S^{\text{ret},\Lambda}(t,t)&=0.\\
    \end{split}
\]
Thus, in this scheme, the Keldysh self-energy does not flow:
\[
    \begin{split}
        \partial_\Lambda \Sigma^{\text{K},\Lambda}(t,t')&=0\\
    \end{split}
\]
 The initial conditions correspond to the self-energy contribution of the thermalized system at infinite temperature:
\[
    \begin{split}
        \Sigma^{\text{K},\Lambda=\infty}(t,t')&=0\\
        \Sigma^{\text{ret},\Lambda=\infty}_{i,j}(t,t')&=\frac{1}{2}\delta(t-t')\sum_{k,l} u_{i,k,j,l}(t).
    \end{split}
\]
Hence, the Keldysh self-energy remains zero throughout the flow and the retarded self-energy is time-local. This type of self-energy has been discussed in Eq.~\eqref{eq:simpelSelf} and Eq.~\eqref{eq:groupProp}. 

Due to the time-local structure of the self-energy, the group property presented in Eq.~\eqref{eq:groupProp} holds and the single-scale propagator of Eq.~\eqref{eq:tdSSsimp} can be simplified further:
\[
    \begin{split}
        S^{\text{ret},\Lambda}(t,t')&=-(t-t')G^{\text{ret},\Lambda}(t,t')\\
        S^{\text{K},\Lambda}(t,t')&=\partial_\Lambda^* \left[G^{\text{ret},\Lambda}(t,0)(1-2\bar n) G^{\text{adv},\Lambda}(0,t')\right]\\
                     &=S^{\text{ret},\Lambda}(t,0)(1-2\bar n) G^{\text{adv},\Lambda}(0,t')\\
                     &\hspace{0.2cm}+G^{\text{ret},\Lambda}(t,0)(1-2\bar n) S^{\text{adv},\Lambda}(0,t')\\
                    &=-(t+t')G^{\text{K},\Lambda}(t,t'),
    \end{split}
\]
resulting in the final form of the flow equation:
\[
    \addtag\label{eq:flowtDep}
    \begin{split}
        \partial_\Lambda \Sigma^{\text{ret},\Lambda}_{i,j}(t,t')&=it\delta(t-t') \sum_{k,l} \left[G^{\text{K},\Lambda}(t,t)\right]_{l,k} u_{i,k,j,l}(t).
    \end{split}
\]
One can show that the causality relation is conserved within this fRG scheme.
By virtue of group property (see Eq.~\eqref{eq:groupProp}), \(G^{\text{K},\Lambda}\) can be computed sequentially~\cite{Kennes2012}:
\[
    \begin{split}
        &G^{\text{K},\Lambda}(t+\Delta, t+\Delta)\\
        &=-iG^{\text{ret},\Lambda}(t+\Delta,0)(1-2\bar n)G^{\text{adv},\Lambda}(0,t+\Delta)\\
        &\hspace{-0.4cm} \stackrel{\text{Eq.~\eqref{eq:groupProp}}}{=}G^{\text{ret},\Lambda}(t+\Delta,t)G^{\text{K},\Lambda}(t,t)G^{\text{adv},\Lambda}(t,t+\Delta).
    \end{split}
\]
Thus, the computationally most expensive part in this scheme is the time evolution (i.e., calculating and applying \(G^{\text{ret}/\text{adv},\Lambda}\)). In the case of a tight-binding chain, one can make use of a Trotter decomposition~\cite{kennes2014universal} of the (in this case tri-diagonal) effective Hamiltonian to speed up the numerics:
For a tri-diagonal matrix \(M\) we define the block diagonal matrices \(M^A\) and \(M^B\) as
\begin{alignat}{2}
    M_{i,i}&=a_i\hspace{1cm} &M_{i,i+1}=b_i=&M_{i+1,i}\notag\\
M_{i,i}^A&=\frac{a_i}{2}\hspace{1cm} &M_{i,i+1}^A=M_{i+1,i}^A&=\left\{\begin{matrix}b_i\hspace{0.5cm} i\ \text{even} \\ 0\hspace{0.5cm} i\ \text{odd} \end{matrix}\right.\\
    M^B&=M-M^A&\notag&
\end{alignat}
such that \(M=M^A+M^B\). The Baker-Campbell-Hausdorff formula then yields
\[
    \begin{split}
        e^M&=e^{\frac{M^A}{2}} e^{M^B}e^{\frac{M^A}{2}}+\mathcal{O}\left(\|M\|^3\right)\\
        &=\lim_{N\rightarrow \infty}\prod_{i=1}^N e^{\frac{M^A}{2N}} e^{\frac{M^B}{N}}e^{\frac{M^A}{2N}}.
    \end{split}
\]
Hence, the error is controlled by the size of \(\|M/N\|\), where \(\|\cdot\|\) denotes a matrix norm. For the applications considered later (i.e., one-dimensional tight-binding models), it is straightforward to reach convergence.

As the exponential function is easily evaluated for this kind of block diagonal matrices, the above provides a computationally efficient way to calculate the retarded and Keldysh Green's function. For large time intervals the most expensive part is to compute the sparse-dense-product \(G^\text{ret}(t+\Delta,t)G^\text{K}(t,t)G^\text{adv}(t,t+\Delta)\), as \(G^\text{K}(t,t)\) will generally be dense for large times, even if the initial \(\bar n\) is sparse.

\section{Excited state fRG}
\label{ssec:exFrg}
We now introduce two ways to obtain correlation functions in excited eigenstates of an interacting Hamiltonian using the fRG, which we label x-fRG. 

Both approaches are based on an adiabatic time-evolution:
The system is first prepared in an excited eigenstate of the noninteracting system and the interaction is then adiabatically turned on. This process is approximated using the functional renormalization group in real-time Keldysh space.

A third method employs the stationary-state Keldysh fRG of Sec.~\ref{ssec:keldFRG} and will be presented as an outlook Sec.~\ref{sec:xFRGw}.
\subsection{x-fRG-\(t\)-\(\Gamma\): Adiabatic time-evolution in Keldysh space with a reservoir cutoff}
\label{ssec:ginfFRG}
\subsubsection{Time-space formulation}
The goal is to use the time-dependent fRG described in Sec.~\ref{ssec:tdfrg} to approximate the evolution of an excited eigenstate of the noninteracting system when interactions are switched on adiabatically.  
Since the self-energy is time-local in this scheme,
\[
    h(t):=h^0+\Sigma^\text{ret}(t)
\]
takes the place of an effective single-particle Hamiltonian. We suppress the \(\Lambda\)-dependence to keep the expressions more readable and introduce the notation
 \[
     \label{eq:gapDef}
     \begin{split}
         &h(t) \left| \Psi_i(t)\right\rangle =\epsilon_i(t) \left| \Psi_i(t)\right\rangle \\
         & h(t)\in \mathbb{C}^{N\times N},\ \hspace{1cm}\left|\Psi_i(t)\right\rangle\in \mathbb{C}^N\\
         &\epsilon_i(t)<\epsilon_j(t) \hspace{2cm} \forall i<j\\
         \bigl|&\epsilon_i(t)-\epsilon_j(t)\bigr|>\Delta \hspace{1cm} \forall i\neq j,t,
     \end{split}
\]
where \(\epsilon_i(t)\) and \(\left|\Psi_i(t)\right\rangle\) denote the instantaneous single-particle eigenvalues and eigenstates of the matrix \(h(t)\). \(\Delta\) is an arbitrary, non-zero gap that ensures the absence of level crossings. Without symmetries, crossings are avoided and such a \(\Delta\) is expected to exist.
The adiabatic theorem~\cite{born1928beweis} then states that if
\[
    \label{eq:adiTheo}
    \begin{split}
        &\hspace{1cm}\| \dot h(t) \| \ll \Delta^2\ \forall t \\
        &\Rightarrow \left| \left\langle \Psi_j(t)\right| \mathcal{U}(t,0) \left|\Psi_i(0)\right\rangle \right|-\delta_{i,j}\ll 1\hspace{0.3cm} \forall i, j
    \end{split}
\]
where \(\mathcal{U}\) denotes the time-evolution operator.
The initial matrix \(\bar n\) associated with a pure many-body eigenstate of the noninteracting, initial Hamiltonian is characterized by a sequence of occupations \(n_i\) of the single-particle eigenstates \(\left| \Psi_i(0)\right\rangle\) of \(h^0\):
\[
    \addtag\label{eq:iniDens}
    \bar n(t=0)=\sum_i n_i \left| \Psi_i(0)\right\rangle \left\langle \Psi_i(0)\right|,\hspace{0.2cm} n_i\in\{0,1\}.
\]
 If the rate of change of the effective Hamiltonian \(h(t)\) is slow enough (compare Eq.~\eqref{eq:adiTheo}), the time evolution of this matrix is given by
\[
    \addtag\label{eq:evolvDens}
    \bar n(t)=\sum_i n_i \left| \Psi_i(t)\right\rangle \left\langle \Psi_i(t)\right|
\]
with the same sequence of \(n_i\). 

As these results are restricted to finite systems they can not directly be applied in presence of an auxiliary reservoir. However, the infinite temperature in the auxiliary reservoirs allows to formally rewrite the Keldysh Green's function (see Eq.~\eqref{eq:simpelSelf}) as
\[
    \addtag\label{eq:adiGK}
    \begin{split}
        G^{\text{K},\Lambda}(t,t)\hspace{-1cm}&\\
        &=-iG^{\text{ret},\Lambda}(t,0)\left[1-2\bar n(t=0)\right]G^{\text{adv}, \Lambda}(0,t)\\
        &=-ie^{-2\Lambda t}\left[1-2G^{\text{ret},\Lambda}_\text{decoup}(t,0)\bar n(t=0)G^{\text{adv}, \Lambda}_\text{decoup}(0,t)\right]\\
    &=:-ie^{-2\Lambda t}\left[1-2\bar n^\Lambda(t)\right],
    \end{split}
\]
where we introduced 
\[
    e^{-\Lambda t}G^{\text{ret},\Lambda}_\text{decoup}(t,t'):=G^{\text{ret},\Lambda}(t,t').
\]
By virtue of this definition \(\bar n^\Lambda(t)\) is the matrix of single-particle correlations of a closed system that evolves according to the effective single-particle Hamiltonian \(h(t)\) (see Eqs.~\eqref{eq:simpelSelf},~\eqref{eq:resSelf} and~\eqref{eq:resG0Decomp}). Thus, \(\bar n^\Lambda(t)\) can be obtained using Eq.~\eqref{eq:evolvDens} and only implicitly depends on \(\Lambda\) (the eigenstates will change with the flow).
Under these circumstances, the flow equation~\eqref{eq:flowtDep} can be rewritten conveniently by the following change of variables:
\[
    \begin{split}
        \xi=e^{-2t\Lambda}&\Rightarrow \partial_\Lambda f(\Lambda)=\Bigl(\partial_\Lambda \xi\Bigr) \partial_\xi f\Bigl(\Lambda(\xi)\Bigr)\\
        \partial_\Lambda&=-2t\xi\partial_{\xi}.
    \end{split}
\]
The final differential equations read:
\vspace{0.5cm}
\\
\underline{\bf x-fRG-\(t\)-\(\Gamma\):}
\[
    \addtag\label{eq:timeXfrgGam}
    \begin{split}
        \partial_\xi \Sigma^{\text{ret},\xi}_{i,j}(t,t')&=-\frac{i}{2}\delta(t-t') \sum_{k,l} \left[\tilde G^{\text{K},\xi}(t,t)\right]_{l,k} u_{i,k,j,l}(t)\\
        \tilde G^{\text{K},\xi}(t,t)&=-i\left[1-2\bar n^\xi(t)\right]\\
        \bar n^\xi(t)&=\sum_i n_i \left|\Psi_i^\xi (t)\right\rangle \left\langle \Psi_i^\xi(t)\right|,
    \end{split}
\]
where \(\xi\) is to be integrated from \(0\) to \(1\).
 Note that this flow equation is simpler than Eq.~\eqref{eq:flowtDep}, since it is to be evaluated only at a single fixed time and interaction.
\subsubsection{Generalizing the fluctuation-dissipation theorem}
\label{sssec:genFDT}
We will now show how Eq.~\eqref{eq:timeXfrgGam} can be evaluated efficiently without the need to fully diagonalize \(h(t)\). 
To this end, we first show how the fluctuation-dissipation theorem in the (effectively) noninteracting case can be applied for nonthermal steady states.
Such states are characterized by a density matrix that commutes with the single-particle Hamiltonian:
\[
    \left[h^0, \bar n\right]=0.
\]
Under this condition there is a common eigenbasis of \(h^0\) and \(\bar n\) in which Eq.~\eqref{eq:thermGtime} immediately leads to
\[
    \begin{split}
        \left[G^\text{K}_0(\omega)\right]_{k,k'}=&-2\pi i(1-2\bar n_{k,k})\delta\left(\omega-\epsilon_k\right)\delta_{k,k'}\\
    =& (1-2\bar n_{k,k})\left[G^\text{ret}_0(\omega)-G^\text{adv}_0(\omega)\right]_{k,k} \delta_{k,k'}.
    \end{split}
\]
If we now assume that \(\bar n_{k,k}\) is already uniquely defined by its energy (which is true, e.g., when the spectrum of \(h^0\) is not degenerate), one can introduce a scalar distribution function
\[
    \bar n_{k,k}=n(\epsilon_k)
\]
such that
\[
    \left[G^\text{K}_0(\omega)\right]_{k,k'}= \left[1-2n(\omega)\right]\left[G^\text{ret}_0(\omega)-G^\text{adv}_0(\omega)\right]_{k,k} \delta_{k,k'},
\]
or in the original basis:
\[
    G^\text{K}_0(\omega)= \left[1-2 n(\omega)\right]\left[G^\text{ret}_0(\omega)-G^\text{adv}_0(\omega)\right].
\]
This is analogous to the fluctuation-dissipation theorem given in Eq.~\eqref{eq:flucDis}. 

Next, we generalize Eq.~\eqref{eq:gkT0}. From now on, we focus on the most relevant case which is a state characterized by the distribution function
\[\addtag\label{eq:distMultiStep}
    \begin{split}
        1-2n(\omega)=\frac{\sigma_{N_\omega}-\sigma_1}{2}+\sum_{i=1}^{N_\omega} \sigma_i \text{sign}(\omega-\omega_i),\\
        \sigma_i\in\{\pm 1\},\ i=1\dots N_\omega 
    \end{split}
\]
where successive \(\sigma_i\) need to have opposite sign and the \(\omega_i\) are in increasing order
\[
    \sigma_i=-\sigma_{i+1},\hspace{1cm} \omega_{i+1}>\omega_{i}
\]
in order to obtain a physical density. Note that the first summand of Eq.~\eqref{eq:distMultiStep} only contributes for an even number of steps. Furthermore, the \(\omega_i\) are assumed not to be part of the spectrum. This object is analogous to the \(n_i\) of the previous section (see Eq.~\eqref{eq:iniDens}) and describes the sequence of occupied and unoccupied single-particle eigenstates. 
As Eq.~\eqref{eq:distMultiStep} can be understood as a superposition of Fermi distributions the considerations of Sec.~\ref{sssec:anaCont} apply for each summand:
\[
    \addtag\label{eq:deformMultiStep}
    G^\text{K}_0(t,t)=i\frac{\sigma_1-\sigma_{N_\omega}}{2}+\sum_j\sigma_j \frac{i}{\pi}\int d\omega e^{-\left|\omega\right|0^+} G^\text{eq}_0(i\omega+\omega_j).
\]

\subsubsection{Frequency-space formulation}
It is now clear that Eq.~\eqref{eq:timeXfrgGam} can be evaluated without the need to fully diagonalize \(h(t)\). Namely, the analytic continuation discussed in Sec.~\ref{sssec:anaCont} immediately yields
\[
    \addtag\label{eq:gkAnalCont}
    \begin{split}
        G^{\text{K},\xi}(t,t)&=i\left(n_1+n_N-1\right)\\
        &\hspace{-1cm}+\sum_{\substack{i\\n_i\neq n_{i+1}}}\frac{i\bar\sigma_i}{\pi}\int d\omega e^{-\left|\omega\right|0^+}\frac{1}{i\omega+\bar\omega_i(t)-h^0-\Sigma^{\text{ret},\xi}(t)}
    \end{split}
\]
where we have introduced
\[
    \begin{split}
        \bar\sigma_i&=n_i-n_{i+1}\\
        \bar\omega_i(t)&=\frac{\epsilon_i(t)+\epsilon_{i+1}(t)}{2}
    \end{split}
\]
in analogy to Eq.~\eqref{eq:distMultiStep}. The \(\epsilon_i(t)\) in this expression are the eigenvalues of the effective single-particle Hamiltonian \(h(t)=h^0+\Sigma^{\text{ret},\xi}(t)\). Hence, Eq.~\eqref{eq:timeXfrgGam} can be evaluated in \(\mathcal{O}(N)\) operations
\footnote{In this, we assume that the sum in Eq.~\eqref{eq:gkAnalCont} only runs over a finite number of changes in \(n_i\) that does not scale with \(N\). 
While there are fast algorithms to obtain the full spectrum of a tri-diagonal matrix in \(\mathcal{O}(N\ln N)\) operations,~\cite{Coakley2013} it is sufficient in this context to compute the eigenvalues at the aforementioned changes in the \(n_i\). This can be done using specialized algorithms like LAPACK's \(dstemr/dstegr\) subroutines; in addition to many other features, these include an implementation of the algorithm presented in Ref.~\onlinecite{Dhillon2004}, which provides orthogonal eigenvectors and eigenvalues at arbitrary positions in the spectrum to working precision in \(\mathcal{O}(N)\) operations.}
for tight-binding models (which we will discuss in Sec.~\ref{sec:compMeth}-\ref{sec:appli}) using Eq.~\eqref{eq:gkAnalCont}. 

However, the functional form of Eq.~\eqref{eq:gkAnalCont} also hints towards a problem: in every step the Green's function is integrated over all imaginary frequencies. This is structurally different from the ground-state flow equation (compare Eq.~\eqref{eq:gsFlow}) and will be investigated more closely in Sec.~\ref{ssec:agreeWithGS}.

The algorithm presented in this section can also be applied to the ground-state but will not necessarily give the same results due to the truncation of the flow equations and difference in cutoff. Hence, comparing the results for the ground-state provides a nontrivial check that will be discussed in Sec.~\ref{ssec:agreeWithGS}.
\subsection{x-fRG-\(t\)-\(\rho_{1,2}\): Adiabatic time-evolution in Keldysh space with an initial-configuration cutoff}
\label{ssec:rhoCut}
We now introduce a second approach which is also based on an adiabatic time-evolution. 
We still use Keldysh space fRG but develop a specialized cutoff for the case that the initial state is an excited eigenstate.

The cutoff in the free Green's functions is introduced by choosing an initial density matrix that depends on \(\Lambda\).
The causality relation is conserved by construction and one can work in the Keldysh basis.
Thus, the cutoff can be introduced on the level of the initial Keldysh Green's function (compare Eq.~\eqref{eq:deformMultiStep}):
\[
    \begin{split}
        G^{\text{K},\Lambda}(0,0)&=G^{\text{K},\Lambda}_0(0,0)=-i(1-2\bar n^\Lambda)\\
        &\hspace{-1.4cm}:=i\left(n_1+n_N-1\right)\\
        &\hspace{-1.25cm}+\sum_{\substack{i\\n_i\neq n_{i+1}}} \frac{i\bar\sigma_i}{\pi}\left[\int_{-\infty}^{-\Lambda}+ \int_{\Lambda}^{\infty}\right] d\omega e^{-\left|\omega\right|0^+}  \frac{1}{i\omega+\bar\omega_i(0) -h^0}
    \end{split}
\]
while the time evolution is still given by Eq.~\eqref{eq:simpelSelf}. This definition of \(\bar n\) differs from that employed in Sec.~\ref{ssec:ginfFRG} and can be understood as a generalization of the sharp cutoff in the Matsubara fRG (compare Sec.~\ref{ssec:gsfrg}). A similar idea has been employed in Ref.~\onlinecite{Jakobs2007a} for steady states of open systems.
By construction, \(\left[h^0,\bar n^\Lambda\right]=0\), and thus the time-evolved Keldysh Green's function \(G^K(t,t)\) can be computed purely from \(h^0+\Sigma(t)\) (compare Sec.~\ref{ssec:ginfFRG}).
The free retarded and advanced Green's functions do not acquire a cutoff in this scheme. The initial condition at \(\Lambda=\infty\) is simply the \(T=\infty\) state while for \(\Lambda=0\) one retrieves the physical density. The time-evolved single-scale propagator in the instantaneous eigenbasis takes the form
\vspace{0.5cm}
\\
\underline{\bf x-fRG-\(t\)-\(\rho_1\):}
\[
    \addtag\label{eq:xfrgrho}
    \begin{split}
        S^{\text{K},\Lambda}(t,t)&=\partial_\Lambda^*G^{\text{K},\Lambda}(t,t)\\
        &\hspace{-1.5cm}=-\hspace{-0.1cm}\sum_k\hspace{-0.1cm}\sum_{\substack{i\\n_i\neq n_{i+1}}}\sum_{\omega=\pm\Lambda}\frac{i\bar \sigma_i}{\pi} \left|\psi_k(t)\right\rangle\left\langle \psi_k(t)\right|\frac{1}{i\omega+\bar\omega_i(0) -\epsilon_k(0)}\\
        S^{\text{ret},\Lambda}(t,t)&=0,
    \end{split}
\]
which together with Eq.~\eqref{eq:genTflow} constitutes the flow equation.
As in the previous scheme, the vanishing retarded component of the single-scale propagator leads to a vanishing flow of the Keldysh self-energy. To calculate the Keldysh component of the single-scale propagator, in general one has to diagonalize the effective Hamiltonian at time \(t\) to obtain the instantaneous eigenstates, which is costly for large systems. A physical approximation can be made by replacing
\[
    \frac{1}{i\omega+\bar\omega_i(0) -\epsilon_k(0)}\rightarrow\frac{1}{i\omega+\bar\omega_i(t)-\epsilon_k(t)}.
\]
 This approximation ensures that the correct number of levels is above and below \(\bar\omega_i(t)\) and only deviates from an exact treatment of Eq.~\eqref{eq:xfrgrho} in higher orders of the interaction since \(\bar\omega_i(0)-\bar\omega_i(t)\in\mathcal{O}(u),\ \epsilon_k(0)-\epsilon_k(t)\in\mathcal{O}(u)\). Using this approximation, \(S^{\text{K},\Lambda}\) takes the more convenient form
\[
    S^{\text{K},\Lambda}(t,t)\approx-\sum_{\substack{i\\n_i\neq n_{i+1}}}\sum_{\omega=\pm \Lambda} \frac{i\bar\sigma_i}{\pi} G^{\text{eq},\Lambda}\left(i\omega+\bar\omega_i(t)\right).
\]
The resulting flow equation reads:
\vspace{0.5cm}\\
\underline{\bf x-fRG-\(t\)-\(\rho_2\):}
\[
    \addtag\label{eq:flowApprox}
    \begin{split}
        \partial_\Lambda \Sigma^{\text{ret},\Lambda}_{n,j}(t,t')&=-\frac{1}{2\pi}\sum_{k,l}\sum_{\substack{i\\n_i\neq n_{i+1}}}\sum_{\omega=\pm\Lambda}\bar\sigma_i u_{n,k,j,l} \\
        &\left[ \frac{1}{i\omega+\bar\omega_i(t) -h^0-\Sigma^{\text{ret},\Lambda}(t)}\right]_{l,k}\delta(t-t'),
    \end{split}
\]
which can be evaluated (for tight-binding chains) in \(\mathcal{O}(N)\) operations.
\section{Comparison of the different schemes}
\label{sec:compMeth}
In this section, we perform various tests to explore the range of validity of the different x-fRG schemes. Due to the first order truncation, results are only guaranteed to agree to linear order in \(u\). All higher orders are uncontrolled and may differ.
\subsection{Tight-binding Hamiltonian}
For the rest of this paper, we focus on a tight-binding model: 
\[
    \addtag\label{eq:tbHamil}
    \begin{split}
    H^\text{tb}=H^\text{hop}+H^\text{int}+H^\text{ph}\\
        H^\text{hop}=\sum_{i=1}^{N-1} c_i^\dagger c_{i+1}+\text{h.c.}\\
        H^\text{int}=U\sum_{i=1}^{N-1} c_i^\dagger c_{i}c_{i+1}^\dagger c_{i+1}\\
        H^\text{ph}=-\frac{U}{2}(c_1^\dagger c_1 +c_N^\dagger c_N)-U\sum_{i=2}^{N-1} c_i^\dagger c_{i},
    \end{split}
\]
where the single-particle index enumerates the Wannier basis states.
Here, \(H^\text{ph}\) is introduced to enforce particle-hole symmetry. We use the hopping amplitude between adjacent sites as the energy-scale.

In the thermodynamic limit, the model described by Eq.~\eqref{eq:tbHamil} is gapless for \(U<2\) and thus its low-energy physics is governed by the Luttinger liquid fixed point.\cite{giamarchi2003quantum} The Luttinger parameter is known from Bethe ansatz calculations and at half filling reads
\[
    \addtag\label{eq:lutParHalf}
    K_{\frac{1}{2}}=\frac{1}{\frac{2}{\pi}\arccos\left(-\frac{U}{2}\right)}=1-\frac{U}{\pi} +\mathcal{O}\left(U^2\right).
\]
For other fillings, the expressions take more involved forms, but for the scope of this paper only the first order expansion is needed at quarter- and three-quarter filling:
\[
    \addtag\label{eq:lutParQuat}
    K_{\frac{1}{4}}=1-\frac{U}{\pi\sqrt{2}}+\mathcal{O}\left(U^2\right)=K_{\frac{3}{4}}.
\]

\subsection{Comparison of the excited state schemes}
Here, we compare the results of the x-fRG-\(t\)-\(\Gamma\) with those obtained from explicit time evolution using the \(t\)-fRG of Sec.~\ref{ssec:tdfrg} as well as the different x-fRG schemes among one another. The x-fRG schemes do not make use of an explicit time-evolution but are instead based on Eq.~\eqref{eq:evolvDens}. Hence, if a gap \(\Delta\) as introduced in Eq.~\eqref{eq:gapDef} exists the x-fRG schemes describe the adiabatic limit of a corresponding scheme with an explicit time-evolution where \(\|\dot h(t)\|\ll \Delta^2\); if higher-order terms are taken into account more care has to be taken. As the x-fRG-\(t\)-\(\Gamma\) and the \(t\)-fRG are based on the same regularization scheme they are expected to yield identical results in this limit while the other schemes are expected to differ in quadratic order in the interaction.

\begin{figure}
    \includegraphics[scale=0.85]{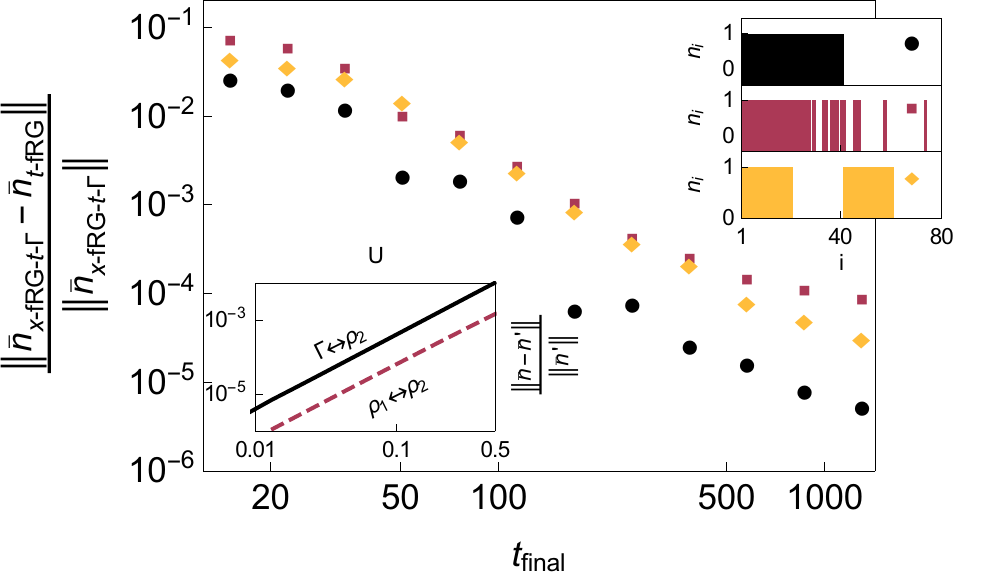}
    \caption{\textit{Main plot:} Relative difference in \(\bar n\) (containing the occupations and all other single-particle correlations, com- pare Eq.~\eqref{eq:rho0}) as predicted by the t-fRG of Sec.~\ref{ssec:tdfrg} and the x-fRG-\(t\)-\(\Gamma\) of Sec.~\ref{ssec:ginfFRG} using the Frobenius norm. For this comparison, a noninteracting tight-binding chain of \(N=80\) sites was prepared in three different many-body eigenstates of the non-interacting system (visualized in the top right inset). Subsequently, the interaction was smoothly increased to its final value of \(U_\text{final}=0.5\) at the final time \(t_\text{final}\). This calculation was repeated for various speeds within the t-fRG and compared with the x-fRG-\(t\)-\(\Gamma\), which is by construction in the adiabatic limit. \textit{Top right inset:} The sequences of \(n_i\) that define the initial states (sorted by energy, see Sec.~\ref{ssec:ginfFRG}). \textit{Bottom left inset:} Deviations as obtained with x-fRG-\(t\)-\(\rho_1\) (dashed line) and the x-fRG-\(t\)-\(\Gamma\) (solid line) relative to the x-fRG-\(t\)-\(\rho_2\) result as function of the final interaction strength \(U_\text{final}\). For this comparison the middle state (red squares in the main plot) was used.}
   \label{fig:comTimeEvolv}
 \end{figure}

To test this, we prepare the system in eigenstates of the noninteracting system with \(N=80\) sites (the three choices used are illustrated in the upper right inset of Fig.~\ref{fig:comTimeEvolv}). The interaction is ramped up smoothly from \(U=0\) to \(U_\text{final}=0.5\) in an increasingly long time span of length \(t_\text{final}\):
\[ 
\frac{U(t)}{U_\text{final}}=\sin^2\left(\frac{t}{t_\text{final}}\frac{\pi}{2}\right) .
\]
The relative difference of the matrix \(\bar n\) containing all single-particle correlations (compare Eq.~\eqref{eq:rho0}) in the final state is shown in Fig.~\ref{fig:comTimeEvolv}.
 The difference between the results decreases with increasing \(t_\text{final}\) with no qualitative difference between the initial states chosen. This explicitly shows that in this case the adiabatic limit can be reached. The saturation at large \(t_\text{final}\) can be explained by numerical inaccuracies in the integration and time evolution. 
 
 The bottom left inset of Fig.~\ref{fig:comTimeEvolv} shows the average difference \(\bar n\) obtained by the x-fRG-\(t\)-\(\Gamma\), x-fRG-\(t\)-\(\rho_1\) and x-fRG-\(t\)-\(\rho_2\) methods. 
 The pairwise difference is quadratic in the interaction, as expected. Even though the x-fRG-\(t\)-\(\rho_2\) and x-fRG-\(t\)-\(\rho_1\) do also differ in quadratic order, their deviation is small compared to the difference between the x-fRG-\(t\)-\(\rho_{1,2}\) and the x-fRG-\(t\)-\(\Gamma\). For this reason, we will not discuss the x-fRG-\(t\)-\(\rho_1\) method in the rest of this paper and instead focus on the computationally cheaper x-fRG-\(t\)-\(\rho_2\). 
\subsection{Comparison with ground-state fRG}
\label{ssec:agreeWithGS}
 Another important test for the different x-fRG-techniques is to compare them to the ground-state fRG of Sec.~\ref{ssec:gsfrg}. As different cutoffs are used and the hierarchy of flow equations is truncated, only agreement to first order is guaranteed.
 
 To study this in detail, we compute the spectral function
\[
    A_i(\omega)=-\frac{1}{\pi} \text{Im} \left[G^\text{ret}(\omega)\right]_{i,i}.
\]
To obtain an approximation of the continuous spectral function of an infinite system, the single-particle levels are broadened by weakly coupling the system to a reservoir. This is equivalent to adding a small imaginary part of the order of the level-spacing to the frequency.
\begin{figure*}
    \includegraphics[scale=0.75]{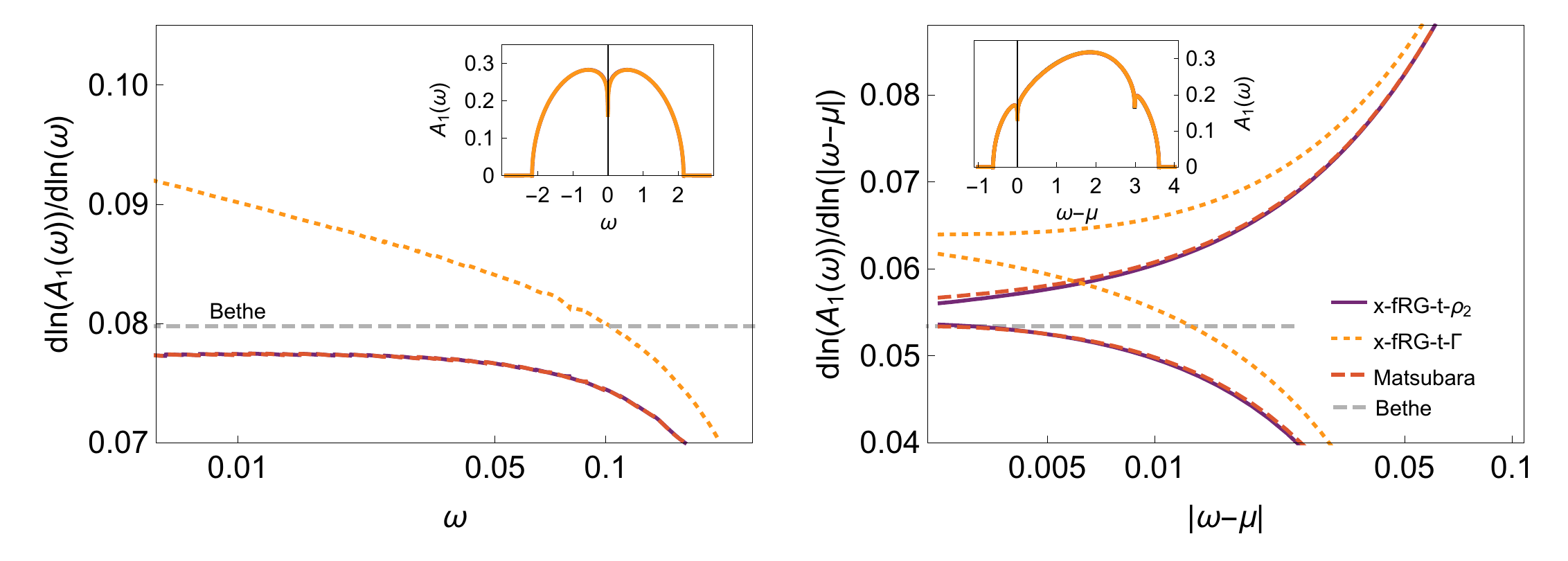}
    \caption{Spectral function at the edge of a tight-binding chain in the ground-state as computed with the ground-state fRG and the excited-state fRG methods (used for the ground-state). The system was chosen particle-hole symmetric with \(N=10^5\) sites and \(U=0.25\) at half and quarter filling (left and right plot, respectively). The insets show the spectral functions while the main plots display the logarithmic derivatives. The horizontal, dashed line shows the Bethe-ansatz expectation for the exponent.}
   \label{fig:comGSlogDeri}
 \end{figure*}

Luttinger liquid theory predicts that for frequencies around the Fermi edge, a power-law suppression occurs in the ground-state at the boundary of our tight-binding chain:
\[
    A_1(\omega)\sim \omega^{\frac{1}{K}-1}.\addtag
    \label{eq:specpower}
\]
The exponent has a nontrivial linear contribution, rendering this spectral function an ideal testbed for the different x-fRG methods.

 For definiteness, we will now discuss the half and quarter filled ground-state. In the half-filled case the approximate x-fRG-\(t\)-\(\rho_2\) scheme (see Sec.~\ref{ssec:rhoCut}) yields the same flow equations as the Matsubara fRG: in the ground-state, there is only one jump in the sequence of occupations of eigenstates (at \(N/2\)). Due to particle-hole symmetry, \(\epsilon_i=\epsilon_{N-i}\) and thus \(\bar \omega_{N/2}(t)=0\). Therefore, Eq.~\eqref{eq:flowApprox} is identical to Eq.~\eqref{eq:flowMasFRG}. In the quarter-filled case, however, Matsubara fRG and x-fRG-\(t\)-\(\rho_2\) deviate.  In contrast, the x-fRG-\(t\)-\(\Gamma\) leads to different flow equations in both, the half and the quarter-filled ground-state.

The spectral function \(A_1\) and its logarithmic derivative are shown in Fig.~\ref{fig:comGSlogDeri}. The logarithmic derivative illustrates that the x-fRG-\(t\)-\(\rho_2\) and the Matsubara fRG indeed yield a power-law suppression. In contrast, the \(T=\infty\) reservoir cutoff used in the x-fRG-\(t\)-\(\Gamma\) fails at this task. This is plausible since the infinite temperature reservoir does not provide a proper low-energy cutoff in the adiabatic limit; all energy scales enter in every step of the calculation (see Eq.~\eqref{eq:gkAnalCont}).

Simple and self-consistent perturbation are known to fail to reproduce power-law behavior in most cases.

\section{Applications}
\label{sec:appli}
\subsection{Generic excitations}
In sections~\ref{sssec:friedel} and~\ref{ssec:cutSpec} we study the survival of Luttinger liquid physics in lowly excited states above the ground-state.
To this end, we choose the initial occupations \(n_i\in\left\{0,1\right\}\) of single-particle eigenstates with respect to the probability distribution  
\[
    \begin{split}
        p_1(n_i=1)&=\frac{1}{1+e^{\left(i-N/2\right)\tilde \beta}}\\
        p_1(n_i=0)&=1-p_1(n_i=1),\\
    \end{split}
\]
where \(\tilde\beta\) acts as an effective inverse temperature in units of the level spacing.
\(\tilde\beta\) should, however, not be understood as a temperature of the interacting state but instead just as a measure of the number of excitations in the system. This scheme is used as a convenient way to obtain excited states in a given energy window. The energy of the final state is however not fixed. 
To make sure the results are independent of the precise way the initial states are generated we also investigate the distributions
\[
    \begin{split}
        p_2(n_i=1)&=\left\{\begin{matrix} 
            1                           &                     &i-N/2&<1/\tilde \beta\\
            \ \frac{1}{2} \hspace{0.2cm}& -1/\tilde \beta \leq&i-N/2&<1/\tilde \beta\\
            0                           & 1/\tilde \beta \leq&i-N/2&
        \end{matrix}\right.\\
        p_3(n_i=1)&=\left\{\begin{matrix} 
            1                                                       &                     &i-N/2&<1/\tilde \beta\\
            \frac{1}{2}-\frac{\tilde\beta (i-N/2)}{2} \hspace{0.1cm}& -1/\tilde \beta \leq&i-N/2&<1/\tilde \beta\\
            0                                                       & 1/\tilde \beta \leq&i-N/2&
        \end{matrix}\right.\\
        p_k(n_i=0)&=1-p_k(n_i=1)\hspace{1cm} k=2,3.
    \end{split}
\]
\begin{figure*}
    \hspace{0.5cm}\includegraphics[scale=0.67]{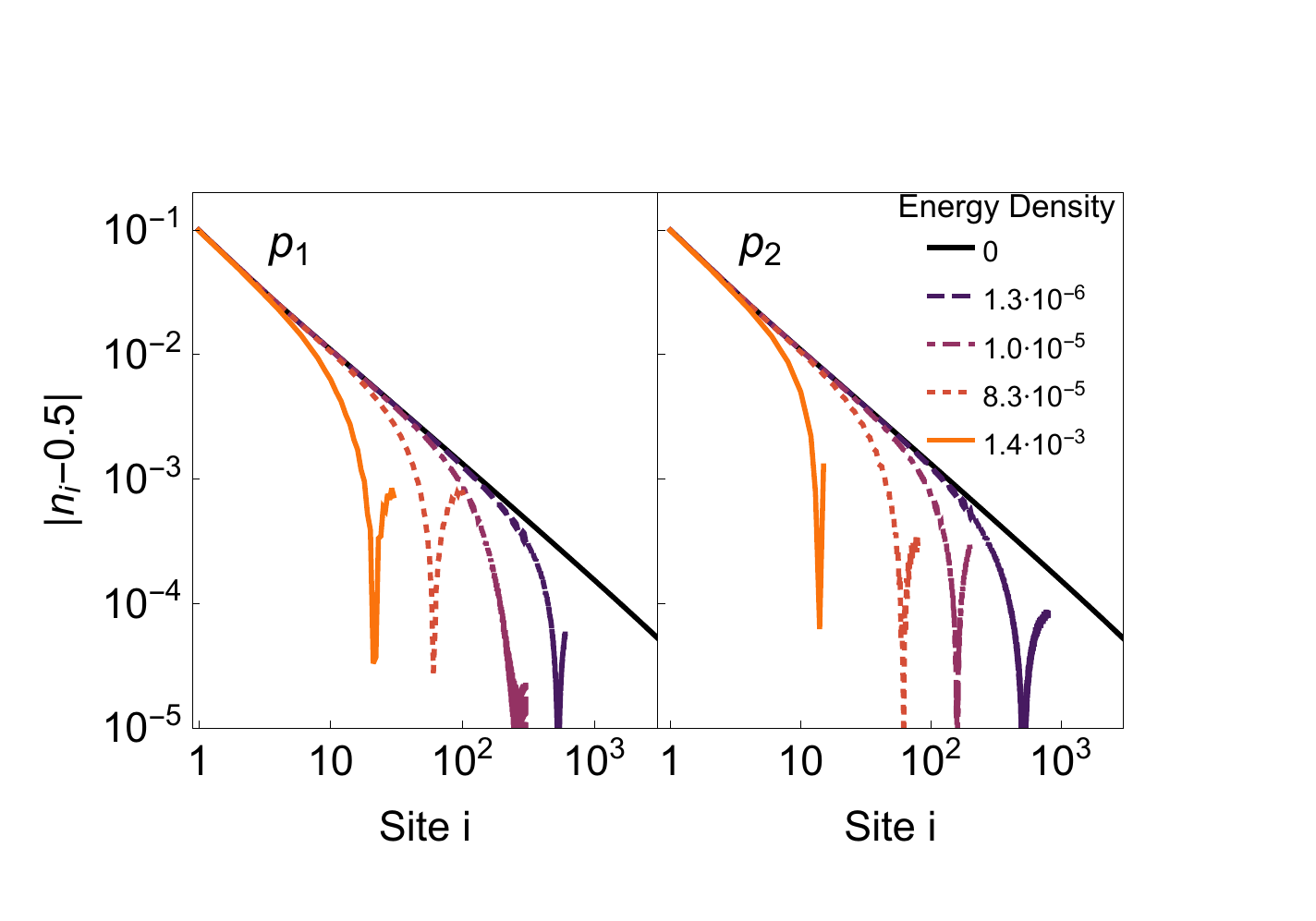}\hspace{-1cm}
    \includegraphics[scale=1.0, trim=0 -0.16cm 0 0]{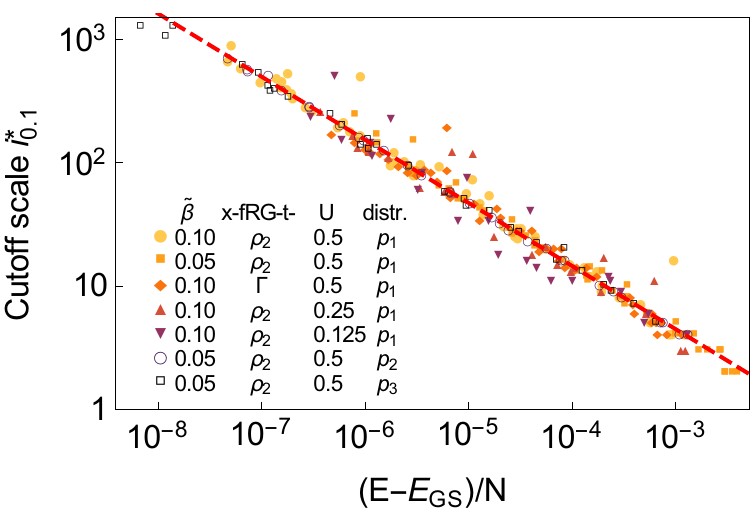}\hspace{0.5cm}
    \caption{\textit{Left plot:} Density profile near the boundary of a tight-binding chain (interaction \(U=0.5\)) prepared in pure eigenstates with different energies above the ground-state. The initial states in the left and right panel were drawn with the probability distributions \(p_1\) and \(p_2\), respectively. The data was obtained via the x-fRG-\(t\)-\(\rho_2\). The Friedel oscillations are cut off on a length-scale related to the energy-density. The system sizes vary between \(N=1.2\cdot 10^3\) and \(N=0.7\cdot 10^5\); for clarity, the curves are only shown up to the cutoff. \textit{Right plot:} Length scale on which the Friedel oscillations are cut off as function of the excitation energy. The interaction varies between \(U=0.125\) and \(U=0.5\), and the size varies from \(N=10^3\) up to \(N=1.6\cdot 10^5\). The behavior is consistent between the probability distributions \(p_{1,2,3}\). The dashed red line is a power-law fit with an exponent \(\alpha\approx0.512\). }
   \label{fig:friedOsc}
    \label{fig:friedCut}
\end{figure*}
The energy of the final state can be obtained computationally cheaply by calculating
\[
    E=\text{Tr}\left[ \bar n \left(h^0+\Sigma\right)\right]
\]
where the \(\bar n\) is only needed on its diagonal and first off-diagonal. Usually, we will be interested in the excitation-energy density defined as \((E-E_\text{GS})/N\).
\subsection{Friedel Oscillations}
\label{sssec:friedel}
We first investigate the Friedel oscillations that emerge around boundaries. A finite tight-binding chain will display oscillations in the density profile (i.e., the occupation numbers) at the ends (or near any other impurity) at zero temperature and if particle-hole symmetry is broken. We restrict ourselves to the case of a clean, half filled, finite chain and set \(H^\text{ph}=0\) (see Eq.~\eqref{eq:tbHamil}) to break particle-hole symmetry explicitly. Luttinger liquid physics predicts that
\[
    \left|\left\langle c_i^\dagger c_i\right\rangle-\frac{1}{2}\right|\sim i^{-K}.
\]
This power law can be obtained via a ground-state fRG calculation, and the linear correction to the non-interacting exponent can be computed from a leading-order scheme.\cite{Andergassen2004}

We now employ our x-fRG framework to explore the effects of finite excitation energies and try to draw analogues with the thermal case: 
At finite temperatures, the algebraic decay is cut off on a length-scale proportional to \(1/T\). 
We investigate whether Friedel oscillations persist in excited eigenstates and if their energy provides a cutoff in a similar fashion as temperature does in the thermal case. 
Since for small $T$ and a linearized dispersion at the Fermi-edge the energy of a thermal state \(E_\text{T}\) scales as 
\[
    E_\text{T}-E_\text{GS}\sim T^2,
\]
we expect the Friedel oscillations to be cut off on a scale proportional to \(1/\sqrt{E-E_\text{GS}}\).

As the x-fRG algorithms scale linear in the number of changes in the sequence \(n_i\) (compare Eqs.~\eqref{eq:gkAnalCont} and~\eqref{eq:flowApprox}), the computational cost can be kept low by fixing the number of single-particle excitations (by choosing a constant \(\tilde\beta\)) while varying the system size. This way, the algorithm scales linearly in the square root of the inverse energy density in the initial state.

The resulting occupation numbers are presented in the left panel of Fig.~\ref{fig:friedOsc} for various states at different excitation-energy densities relative to the ground-state. One can see that even at finite energy density a power-law decay is visible close to the boundary. This decay is cut off at a energy dependent length-scale. To objectively measure the deviations from the ground-state Friedel oscillations, we define a cutoff scale
\[
    i^*_\alpha=\min\left\{i\middle|\frac{|n_i^\text{GS}-n_i^\text{ex}|}{|n_i^\text{GS}-0.5|}>\alpha\right\}
\]
such that \(i^*_\alpha\) gives the first site where the relative error compared to the Friedel oscillations of the ground-state exceeds \(\alpha\). This scale is plotted in Fig.~\ref{fig:friedCut} as a function of the excitation energy density for various eigenstates and systems of different lengths. The line is a power-law fit resulting in 
 
\[i^*_{0.1}\sim \left(\frac{E-E_\text{GS}}{N}\right)^{-0.512},\]
which is reasonably close to the thermal expectation of \(-1/2\).
This result does not depend on the chosen \(\bar\beta\), the cutoff or the distribution used to generate the initial state.
We thus conclude that Friedel oscillations survive to finite energy-densities and that the excitation energy density simply provides an infrared cutoff.

Due to the shortcomings of x-fRG-\(t\)-\(\Gamma\) discussed in Sec.~\ref{sec:compMeth} we solely employ the x-fRG-\(t\)-\(\rho_2\) from now on.
\subsection{Spectral function at \(\omega=0\)}
\label{ssec:cutSpec}
Another characteristic of Luttinger liquids is the finite-temperature behavior of the local spectral function introduced in Eq.~\eqref{eq:specpower} at \(\omega=0\). In a thermal state
\[
    A_1(\omega=0)\sim T^{\frac{1}{K}-1}.
\]
In analogy to the previous section we thus investigate whether in an generic, excited many-body eigenstate
\[
    A_1(\omega=0)\stackrel{?}{\sim} \left(\sqrt{\frac{E-E_\text{GS}}{N}}\right)^{\frac{1}{K}-1}.
\]

\begin{figure}
    \includegraphics[scale=0.75]{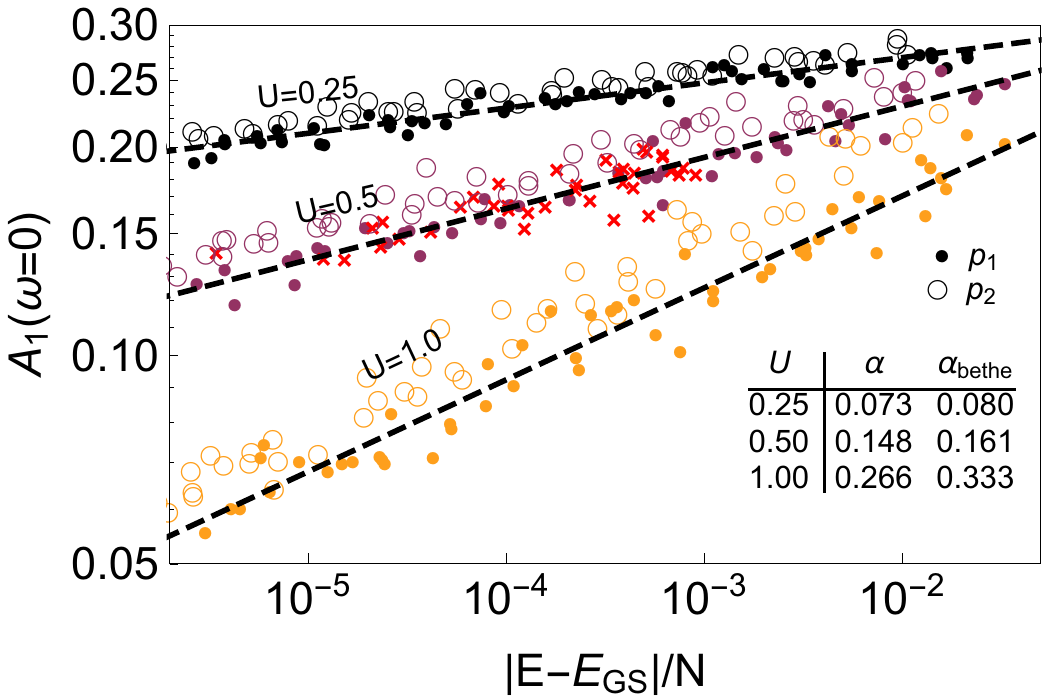}
    \caption{Local spectral function on the first site of a chain in an excited eigenstate. The circles show systems of lengths varying form \(N=4\cdot 10^2\) to \(N=5\cdot 10^4\) using the probability distribution \(p_1\) (filled circles) and \(p_2\) (empty circles) with \(\tilde \beta=0.05\) at the indicated interaction strengths. To obtain the data represented by the red crosses a fixed system size \(N=5000\) was used varying \(\tilde\beta\) from \(0.025\) to \(0.3\) in the distribution \(p_1\) with a fixed \(\gamma=0.0025\). The results obtained at fixed system size and \(\gamma\) are compatible with those obtained from simultaneously varying the size and Lorentzian broadening justifying this procedure.} 
   \label{fig:specWithExci}
 \end{figure}
As discussed before, it is computational advantageous to keep \(\tilde \beta\) constant while varying the system-size. To compensate for finite-size effects a small imaginary part \(\gamma\) is added to the frequency when evaluating the spectral function, effectively widening the sharp single-particle levels to Lorentzian peaks. We chose \(\gamma=1/(N\tilde \beta)\). 
To ensure that \(\gamma\) is big enough to remove finite-size effects while being small enough not to influence the observed behavior
Fig.~\ref{fig:specWithExci} also shows the results obtained at a fixed size \(N=5000\) and fixed smearing \(\gamma=2.5\cdot 10^{-3}\) by varying \(\tilde \beta\) of the distribution \(p_1\). As the data obtained in either way is compatible, the scaling observed can be attributed to the excitation energy. Data was obtained for states generated with the distribution \(p_1\) and \(p_2\). While the data shows different prefactors, the exponents are similar. For clarity, the analysis of the exponent will thus be restricted to the distribution \(p_1\). The exponent is extracted with a power-law of the form 
\[A_1(i\gamma)\sim \left(\sqrt{\frac{E-E_\text{GS}}{N}}\right)^{\alpha}
\]
and compared to the thermal expectation from Bethe-ansatz calculations (see Fig.~\ref{fig:specWithExci}).
 
 The results indicate that generic excitations can not only provide a cutoff but can also be the origin of Luttinger liquid-like power-laws.

\subsection{Block excitations}
\label{ssec:block}
\begin{figure*}
    \includegraphics[scale=0.7]{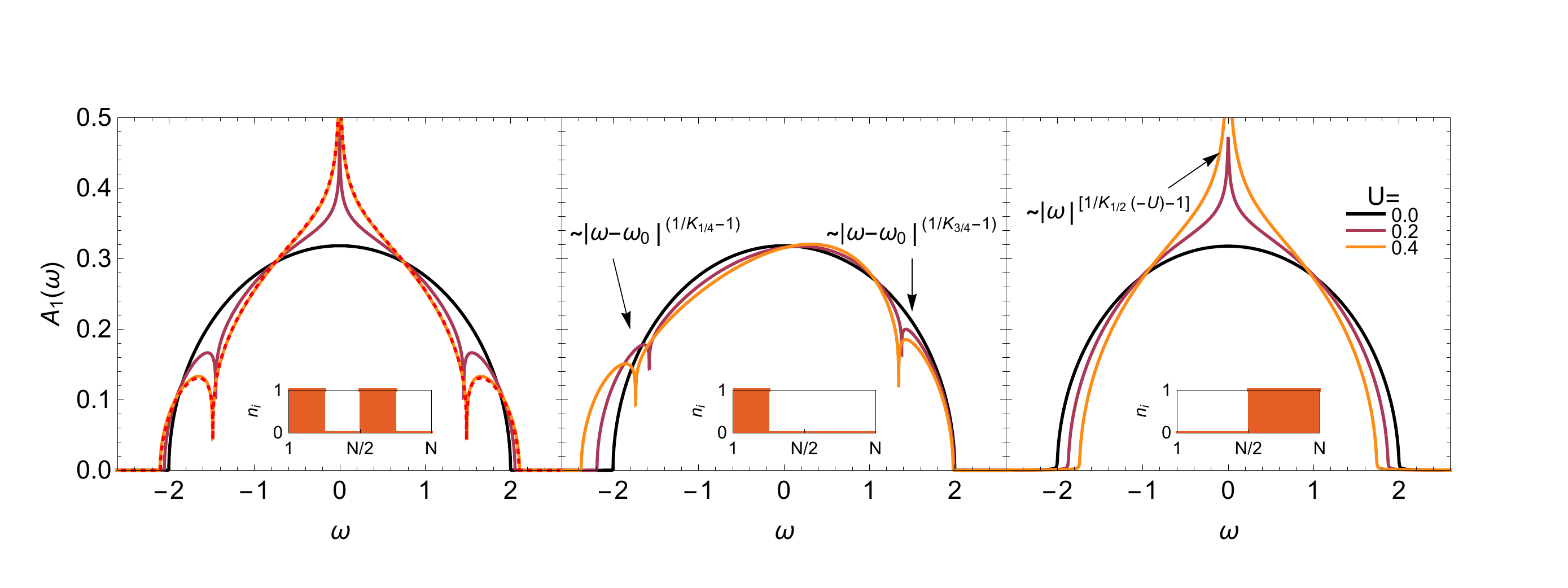}
    \caption{Spectral function of a tight-binding chain at the boundary. The interactions are \(U=0,0.2,0.4\) and the initial sequence of occupied eigenstates of the noninteracting system is shown in the insets. The results shown as solid lines were obtained using the x-fRG-\(t\)-\(\rho_2\). \textit{Left panel:} Block excitation defined by Eq.~\eqref{eq:blockEx} in a system of \(N=10^5\) sites. \textit{Center panel:} Quarter filled ground-state with \(N=10^5\). The three-quarter filled ground-state gives the mirrored result. \textit{Right panel:} Inverted ground-state using \(N=4\cdot10^3\) sites. The dashed, red line represents results obtained using the x-fRG-\(\omega\) presented in Sec.~\ref{sec:xFRGw}.}
   \label{fig:dosPanel}
 \end{figure*}

\begin{figure}
    \includegraphics[scale=1.00]{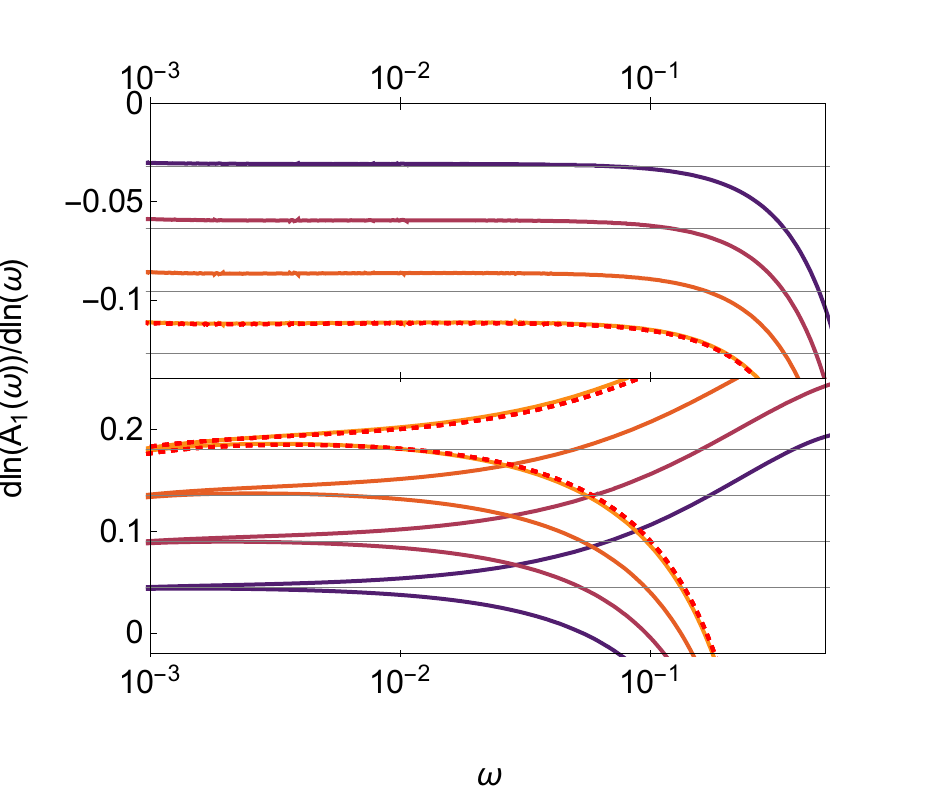}
    \caption{Logarithmic derivatives of the power-laws in the spectral function at the boundary in a block-excited state of a tight-binding chain (compare left panel of Fig.~\ref{fig:dosPanel}). The top panel shows the derivative around \(\omega=0\) while the bottom panel shows the two sides around the outer cusp. The interactions depicted are (from dark to light colors) \(U=0.1,0.2,0.3,0.4\). The horizontal lines in the respective panels show \(-U/\pi\) and \(\sqrt{2}U/\pi\). The dashed, red line represents results obtained using the x-fRG-\(\omega\) presented in Sec.~\ref{sec:xFRGw}.}
   \label{fig:dosDeri}
 \end{figure}

So far, we have analyzed generic excitations which leave the general monotonic shape of the function \(n(\omega)\) unchanged and just alter it around the Fermi-edge where the (free) dispersion is nearly linear.
As all observables in a Luttinger liquid are governed by the Hamiltonian at low energies, it is not surprising that the results are similar to those of thermal states. There is, however, a different set of excited states which are highly non-generic and different from thermal states, and thus we have no intuition for their physics. As an example, we consider a state where a whole block of fermions is excited to a higher energy region. In the simplest, most symmetric case this is modeled by
\[
    \addtag\label{eq:blockEx}
n_i=\left\{\begin{matrix}
&1\hspace{1cm}& &i\leq&\frac{1}{4}N\\
&0\hspace{1cm}&\frac{1}{4}N<&i\leq&\frac{1}{2}N\\
&1\hspace{1cm}&\frac{1}{2}N<&i\leq&\frac{3}{4}N\\
&0\hspace{1cm}&\frac{3}{4}N<&i\leq&N
\end{matrix}
 \right.
\]
which in contrast to the ground-state has three sharp, distinct edges in the distribution function.
To investigate the physics in this state, we again study the spectral function at the boundary (compare Sec.~\ref{ssec:agreeWithGS}, we employ the same Lorentzian broadening). Results are shown in the first panel of Fig.~\ref{fig:dosPanel}, where one can identify several points of non-analytic behavior.
\begin{figure}
    \includegraphics[scale=.7]{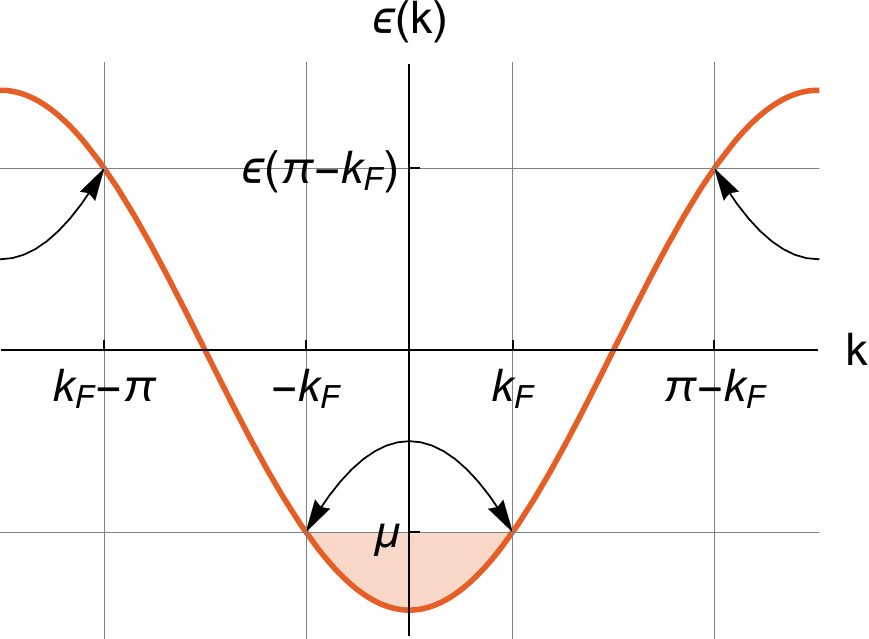}
    \caption{Illustration of umklapp processes contributing a second cusp in the local density of states for all filling fractions other than the half filled case (compare center panel of Fig.~\ref{fig:dosPanel}).}
   \label{fig:umklapp}
 \end{figure}%
The logarithmic derivatives around these points are plotted in Fig.~\ref{fig:dosDeri} and indicate the existence of power-laws. The spectral function around the center divergence seems to scale as
\[
    A_1(\omega)\sim\left|\omega\right|^{-\frac{U}{\pi}}
\]
while the outer cusps are described by
\[
        A_1(\omega)\sim\left|\omega-\omega_0^{1,3}\right|^{\frac{U\sqrt{2}}{\pi}}.
\]
The rest of this section is devoted to understanding this result in terms of known Luttinger liquid physics. 

To this end, the center panel of Fig.~\ref{fig:dosPanel} shows the spectral function in the quarter filled ground-state as obtained with the x-fRG-\(t\)-\(\rho_2\); the three-quarter filled case is the same but mirrored. Both show two cusps, one at the Fermi energy and a second one at the energy associated with \(\pi-k_F\). The first cusp is the one well studied in the literature while the second cusp is the result of umklapp scattering~\cite{AndergassenPHD}: the long-ranged potential, acquired by the Friedel oscillations, allows for processes with wave vector \(2k_F\), which is either close to the Fermi surface or an umklapp term higher up in the spectrum (compare Fig.~\ref{fig:umklapp}). 
Each of these is described by a power-law \(A_1(\omega)\sim\left|\omega-\omega_0\right|^{1/K-1}\). \(K\) has to be calculated using the dispersion at the corresponding position in the spectrum (i.e., at quarter and three-quarter filling), and to first order the exponent is given by \(U/\pi\sqrt{2}\) (compare Eq.~\eqref{eq:lutParQuat}).

Next, we discuss the inverted half filled ground-state (right panel of Fig.~\ref{fig:dosPanel}) which can still be obtained from the ground-state algorithm by using a infinitesimal negative temperature, or equivalently by analyzing the ground-state of \(-H\). As the sign of the hopping is arbitrary, we are thus just inspecting the ground-state of a tight-binding chain with attractive interactions; to leading order, the corresponding exponent is just \(-U/\pi\).

We now observe that the left panel of Fig.~\ref{fig:dosPanel} can be interpreted as a superposition of the effects associated with the different jumps in the distribution function: The first and third discontinuity are located at the Fermi energy of a quarter- and three-quarter filled system, respectively. Their contributions coincide and (in the present approximation scheme) the exponents just add up to 
\[2\left(\frac{1}{K_\frac{1}{4}}-1\right)=\frac{U\sqrt{2}}{\pi}+\mathcal{O}\left(U^2\right)=2\left(\frac{1}{K_\frac{3}{4}}-1\right).\]
The divergence in the center is described purely by the inverted ground-state.

Hence, the phenomenology of the very highly excited, non-generic case of a block-excitation can (at least to leading order) be interpreted in terms of ground-state Luttinger liquid physics.
\section{Towards second order: Steady-state Keldysh fRG with a non-thermal reservoir cutoff}
\label{sec:xFRGw}
We now devise a third way of obtaining an effective noninteracting description of an excited eigenstate for an interacting model which is not based on an adiabatic time-evolution but instead employs the steady-state Keldysh formalism of Sec.~\ref{ssec:keldFRG}. 
This could be of central importance in going beyond linear order in interactions: the x-fRG-\(t\)-\(\Gamma,\rho\) discussed before are based on a picture of discrete, separated levels. In second order these levels will be broadened resulting in a finite lifetime of quasiparticles. Hence, a picture of adiabatic time-evolution becomes questionable. While a generalization to second order of the algorithm presented here is still non-trivial it could be a promising starting point. Whether a second-order treatment would result in relevant entanglement to the auxiliary bath at arbitrarily weak couplings remains to be investigated.

The system is assumed to be in the steady-state induced by a wide-band reservoir. As in Sec.~\ref{ssec:keldFRG}, the hybridization to this reservoir is used as the cutoff, but the reservoir is now no longer in equilibrium but instead chosen to be in a non-thermal, pure state described by the distribution function (compare Sec.~\ref{sssec:genFDT})
\[
    \label{eq:distResCut}
        1-2n(\omega)=\frac{\sigma_{N_\omega}-\sigma_1}{2}+\sum_{i=1}^{N_\omega} \sigma_i \text{sign}(\omega-\omega_i),\\
    \] 
At the beginning of the flow where the coupling \(\Lambda\) to the reservoir is strong, all levels are infinitely broadened and thus empty, half-filled or full (according to the first term of eq.~\eqref{eq:distResCut})  and uncorrelated (i.e., described by a \(T=\infty\) state, independently of the precise state of the reservoir). For small couplings (i.e., at the end of the flow), the energy-scales of the physical Hamiltonian dominate and the distribution function governing the steady state becomes equal to the one of the reservoir. Hence, one recovers the physical system featuring a non-thermal distribution function.

This cutoff procedure leads to a flow equation of the form (see Eq.~\eqref{eq:flowKeldFRG})
\vspace{0.5cm}\\
\underline{\bf x-fRG-\(\omega\):}
\[
    \begin{split}
        \partial_\Lambda \Sigma^{\text{ret},\Lambda}_{n,m}&\\
        &\hspace{-1.2cm}=-\frac{1}{2\pi}\sum_{k,l}\sum_i\sigma_i u_{n,k,m,l} \left[G^{\text{ret},\Lambda}(\omega_i)+G^{\text{adv},\Lambda}(\omega_i)\right]_{l,k}\\
        &\hspace{-1.2cm}=-\frac{1}{2\pi}\sum_{k,l,i}\sum_{\omega=\pm\Lambda}\sigma_i u_{n,k,m,l} \left[ \frac{1}{i\omega+\omega_i -h^0-\Sigma^{\text{ret},\Lambda}}\right]_{l,k},
    \end{split}
\]
which is remarkably similar to the one found in the schemes discussed previously (see Eq.~\eqref{eq:flowApprox}) and even coincides, if the relevant parts of the spectrum are invariant when lowering the cutoff.
In general, however, the effective single-particle spectrum will change during the flow. By definition of the cutoff scheme, at no point the particle number is fixed, only the occupation in energy space. The actual number of particles in the final system cannot be fixed beforehand. To obtain a state with a desired sequence of occupations \(n_i\) of the effective single-particle levels the frequencies \(\omega_i\) have to be iteratively adjusted. 

For the case of a block excitation this optimization procedure is straightforward and the results of such a calculation are shown as dashed lines in Fig.~\ref{fig:dosPanel}. For this problem, the x-fRG-\(\omega\) produces power-laws; to obtain more generic states using this method is, however, connected to a significant overhead.

 \section{Summary and outlook}
In this paper, we have shown how correlation functions in pure excited states of many-body systems can be
obtained within the realm of the functional renormalization group; the key idea is
to start out with a Slater determinant and to slowly switch on interactions. To this end, we simplified existing
real-time Keldysh fRG flow equations for the special case of adiabaticity (x-fRG-$t$-$\Gamma$) and also
devised a novel cutoff scheme which is specifically tailored to this problem
(x-fRG-$t$-$\rho$). Due to the approximate nature of the method, the x-fRG results do not
necessarily agree with those of standard equilibrium fRG when targeting the ground
state, which thus provides a nontrivial testing ground. Importantly, only the
x-fRG-$t$-$\rho$ manages to reproduce the power-law suppression of the spectral function at the
boundary of a Luttinger liquid; the x-fRG-$t$-$\Gamma$ fails at this task.

We subsequently employed the x-fRG to study two toy problems. First, we
demonstrated that Luttinger liquid power law behavior survives in lowly-excited pure
states whose excitation energy density serves as an infrared cutoff. Second, we determined the spectral function of highly-excited, nongeneric block excitations featuring multiple Fermi edges and illustrated that the system
is effectively governed by a superposition of several Luttinger liquids.

The key drawback of the x-fRG is its approximate character. Even though the
underlying RG idea entails an infinite resummation of Feynman diagrams, all results
presented in this paper are only guaranteed to be correct up to leading order in the
interaction. The strengths of the x-fRG are that it is not bound by the growth of
entanglement and that large systems of up to $10^6$ sites (in one dimension) can be
treated easily.

Future directions include an extension of the x-fRG flow equations to second order. The x-fRG-\(\omega\) could provide a good starting point for a second-order treatment as it is less dependent on an effective single-particle picture. This cutoff procedure can be readily applied in second-order opening up the interesting question of whether the final state remains pure up to the truncation order.

 The methodology developed in this paper is directly applicable to question arising in the field of many-body localization.\cite{Gornyi2005,Basko2006} Using this one can obtain access to the entire spectrum of an interacting disorder system. In future work we hope to contribute to the debate about the existence of a mobility edge separating localized and delocalized eigenstates.

Going beyond this, an application to many-body localized, topological systems is enticing.
Because of the small overlap of far apart localized states, explicit adiabatic time-evolution is expected to be exponentially hard. Due to the smooth distribution function used in the flow-equations of the x-fRG-\(t\)-\(\rho\) it might be a promising method to circumvent this problem and explore the many-body spectrum of interacting, localized systems.

\section*{Acknowledgements}
We thank Volker Meden for comments on our manuscript and acknowledge support by the Deutsche Forschungsgemeinschaft through the Emmy Noether program (Grant No. KA 3360/2-1) and the CRC/Transregio 183 (Project A03). DMK was supported by the Basic  Energy Sciences Program of the U. S. 
Department of Energy under Grant No. SC-0012375 and by DFG KE 2115/1-1.
DMK also acknowledges the hospitality of the Center for Computational Quantum Physics of the Flatiron Institute.

\bibliography{./references_NoArx,./referencesThes}

\end{document}